\documentclass{aa}  

\usepackage{graphicx}
\usepackage{txfonts}
\usepackage{xcolor}

\usepackage{hyperref}
\hypersetup{
    colorlinks=true,
    linkcolor=blue,
    citecolor = blue,
    urlcolor=black
    }
\begin{document}
   \title{Shear, writhe and filaments: turbulence in the high latitude molecular cloud MBM 40}

\author{Marco Monaci
          \inst{1}
          \and
          Loris Magnani
          \inst{2}
          \and
          Steven N.Shore
          \inst{1,4}
          \and
          Henrik Olofsson
          \inst{3}
          \and
          Mackenzie R. Joy
          \inst{2}
          }

\institute{Dipartimento di Fisica, Università di Pisa, Largo Bruno Pontecorvo 3, Pisa \\
              \email{monaci93@gmail.com} \\
            \email{steven.neil.shore@unipi.it}
         \and
             Department of Physics and Astronomy, University of Georgia, Athens, GA 30602-2451 \\
             \email{loris@uga.edu} \\
             \email{mackenzierosejoy@gmail.com}
         \and
              Department of Space, Earth and Environment, Chalmers University of Technology, Onsala Space Observatory, 43992 Onsala, Sweden \\
             \email{henrik.olofsson@chalmers.se}
         \and
             INAF-OATS, Via G.B. Tiepolo 11, 34143 Trieste, Italy}
             

\date{Received: -; accepted: - }

  \abstract
   {It is almost banal to say that the interstellar medium (ISM) is structurally and thermodynamically complex. But the variety of the governing processes, including stellar feedback, renders the investigation challenging. High latitude molecular clouds (HLMCs) with no evidence of internal star formation, such as MBM 40, are excellent sites for studying the chemistry and dynamic evolution of the cold neutral ISM.}
   {We used this high latitude cloud as an exemplar for the dynamical and chemical processes in the diffuse interstellar medium.}
   {We analyzed new and archival \element[][12]CO, \element[][13]CO, CH, HCO$^+$, CS, H$_2$CO, HCN data from Five College Radio Observatory (FCRAO), Onsala Space Observatory (OSO), Arizona Radio Observatory (ARO) and W. Gordon telescope (Arecibo) combined with the Galactic Arecibo L-band Feed Array \ion{H}{i} (GALFA-HI) \ion{H}{i} 21 cm data set, to study the chemistry, thermal state, and dynamics of MBM 40. A new dynamical analytical  approach was adopted by considering each line profile as a line of sight Probability Distribution Function (PDF) of the turbulence weighted by gas emissivity.}
   {The atomic and molecular gas are smoothly distributed in space and velocity. 
   No steep transition is seen between circumcloud atomic and cloud molecular gas in either radial velocity or structure. We proposed a topology of the cloud from the molecular tracers, a  contorted filamentary structure that is shaped by a broad embedding shear flow in the neutral atomic gas.  Comparative  examination of different molecular tracers shows that \element[][13]CO, H$_2$CO and CS arise from only denser molecular cores, where \element[][12]CO, CH and HCO$^+$ traces diffuse gas with broader range of dynamics.}

  
\maketitle

\section{Introduction: The turbulent context}

Molecular lines routinely detected in diffuse interstellar clouds  (e.g. \cite{2005IAUS..231..187L})  by centimeter and millimeter radio transitions present motions on a broad range of lengths and velocities (\cite{2011piim.book.....D}, \cite{2014NPGeo..21..587F}, \cite{1999ApJ...512..761L}, \cite{2003ApJ...593..413S}, \cite{2006A&A...457..197S}).  It is almost a commonplace to say that understanding the dynamics presents both observational and theoretical challenges.  But the reasons derive, in part, from the extension of laboratory analytical methods when trying to generalize from experiments to astronomical observations.  

Several techniques have been introduced based on laboratory analogies to study the dynamical structure and characterize the flows in molecular clouds. \cite{2005ApJ...631..320E} and \cite{2008A&A...481..367H} used velocity centroid maps to study the turbulence and the intermittency visible in profiles; principal component analysis was investigated by, for example, \cite{2013MNRAS.433..117B}, and \cite{2000ApJ...537..720L} used Fourier analysis of two-dimensional images integrated  in both narrow and wide velocity ranges. \cite{1994ApJ...429..645M} computed structure functions (SF) from centroid velocity maps and found a scaling exponent for second-order SF in the range of 0.4 - 1.4 with a mean value of 0.86. Virtually all molecular clouds show a similar behavior, implying the universality of turbulence. 

In astrophysical studies, identifying the source for the driving of the flow dynamics is complicated by something that is not an issue in terrestrial settings.  The driving can be local, such as shocks, stellar winds or explosions on sub-parsec scales, or global, such as Galactic rotation or density waves. Moreover, the dissipation scale that defines the bottom of the energy cascade (e.g., \cite{tennekes1972first}, \cite{2000tufl.book.....P}, \cite{1990cp...book.....M}) is still difficult to observe and even more difficult to model.  Finally, the fundamental difficulty presented by interstellar turbulence is that observations provide only an instantaneous snapshot of only one velocity component -- radial velocities -- with poor spatial resolution (whatever the overall spatial coverage is of any survey).  

Observing high latitude molecular clouds (\cite{2015ARA&A..53..583H}, \cite{2017ASSL..442.....M}) may render the intrinsic nature of these flows more accessible (see \cite{2004ARA&A..42..211E}, \cite{2012A&ARv..20...55H} for a generic perspective), because, in most cases, they show no evidence of internal star formation (\cite{1999A&A...341..163H}, \cite{1996ApJ...465..825M}).  
\cite{1990ApJ...359..344F} found  that the velocity fields in non-star forming regions show a Kolmogorov-like spatial scaling of the velocity dispersion and proposed that the large deviation from Gaussian wings in the line profiles is a signature of intermittency, a fundamental property of turbulent motion (see, for instance, \cite{2000tufl.book.....P} or \cite{2008tufl.book.....L}). \cite{1991ApJ...378..186F} found that maps at different spatial scales are self-similar, possibly fractal, thus consistent with turbulent dynamics inside molecular clouds. 

Besides non-thermal linewidths, several molecular complexes also present  systematic  velocity gradients on a parsec scale,  observed in atomic and molecular gas (see  \cite{2011ApJ...732...78I}). The physical process(es) that produce  these are not fully understood. Imara and Blitz proposed cloud rotation, where the change in the angle between the line of sight and velocity mimic the velocity gradient; \cite{2006ApJ...638..191K} proposed large-scale motions as driving mechanism for the observed velocity gradients. Such ordered motions are evident, as we will show, in MBM 40, in addition to turbulent broadening, providing a striking example of a cloud that mimics some key aspects of laboratory flows.

\subsection{Introducing MBM 40}
MBM 40 is a low mass, diffuse\footnote{\cite{1998ApJ...500..525S} indicate a maximum E(B-V) of 0.24 mag for MBM 40. Using the standard relation between A$_V$ and E(B-V), the maximum A$_V$ would be about 0.7-0.8 mag, making MBM 40 a diffuse rather than a translucent cloud according to the definition proposed by \cite{1988ApJ...334..771V}. Furthermore, from \ion{H}{i} calibration we have an independent measurement of A$_V \sim$ 0.5, compatible with the SFD dust map.}, high-latitude ($\ell \sim35^\circ$; $b \sim+45^\circ$), non-star forming, molecular cloud (\cite{1996ApJ...465..825M}) embedded in a larger \ion{H}{i} structure as highlighted by \cite{2003ApJ...593..413S}. The cloud has been extensively studied in past years (\cite{1985ApJ...295..402M}, \cite{2002JKAS...35...97L}, \cite{2006A&A...457..197S}, \cite{2010AJ....139..267C}, \cite{2012AJ....144..163C}, \cite{2013MNRAS.436.1152C}). The cloud mass, inferred by \element[][12]{CO}{} observations, is 20 - 40 M$_\odot$. The absence of star formation and other internal processes, such as shocks, presents a cleaner example of the turbulent dynamics and  cold chemistry. At a distance of 93$^{+23}_{-20}$ pc (\cite{zucker2019}), MBM 40 is one of the closest molecular clouds, allowing sampling on a 
spatial resolution of around 10$^{-2}$ pc.
The cloud shows dense substructures with $n_{\mathrm{H}_2} \geq 10^3$ cm$^{-3}$ and N(H$_2$) $\sim$ (3 - 8) $\times$ 10$^{20}$ cm$^{-2}$, N(\ion{H}{i}) $\sim$ (1 - 2) $\times$ 10$^{20}$ cm$^{-2}$. MBM 40 is well isolated in velocity with \element[][12]CO observations confined within a velocity range [+2.5,+4.2] km s$^{-1}$ (\cite{2022A&A...668L...9M}). Large CO maps and infrared images  show a hairpin structure, but the infrared emission is  more extended than the molecular gas. In the present work we focus on the western ridge of the cloud that exhibits a higher density than the eastern side.

\section{Description of the observational data}\label{sec:observations}
\begin{figure*}
  \centering
  \includegraphics[scale=0.35,bb=0 0 1450 700,clip]{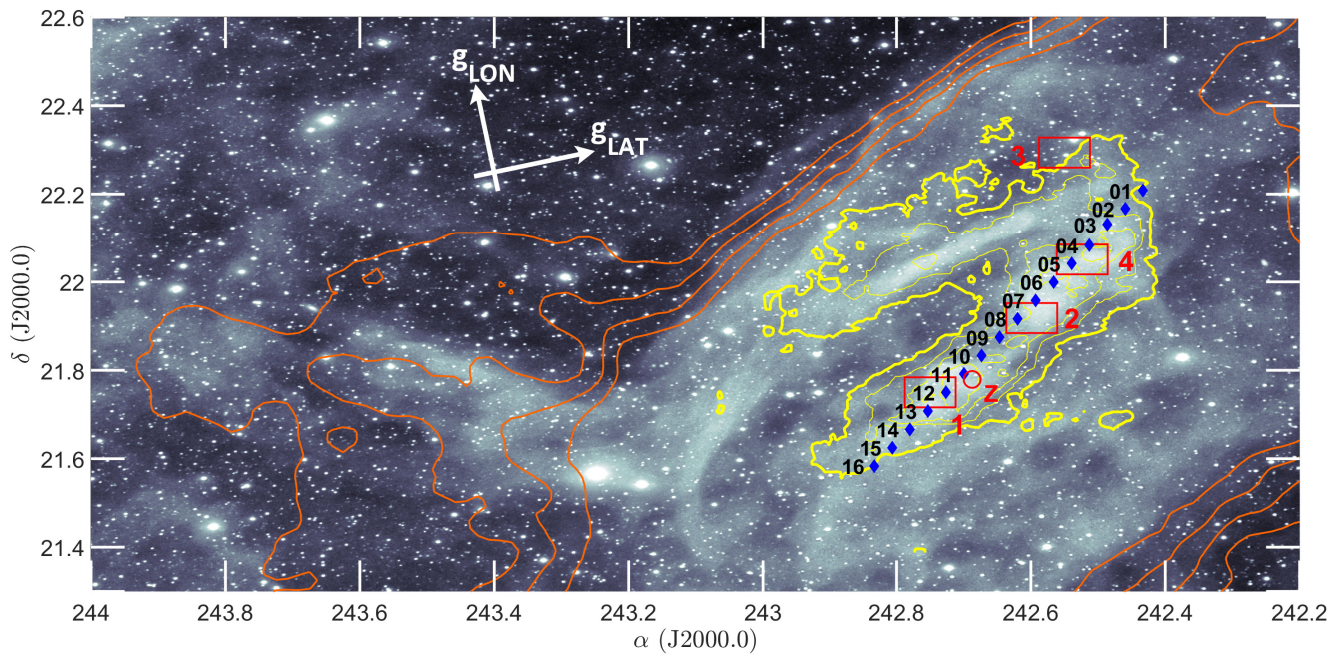}
     \caption{Ensemble of observed positions for MBM 40. The molecular morphology derived from FCRAO \element[][12]CO contours at 1, 2, 3, 4 K km $\mathrm{s}^{-1}$ is shown in yellow. Red rectangles with numbers are MAP1, 2, 3 and 4 observed mainly in \element[][12]CO and \element[][13]CO with Onsala telescope. Blue diamonds with numbers along the west ridge are the Arecibo CH positions used in the shear analysis in \S 5.2 (other archival CH observations, which cover the whole cloud, are not shown here). Position Z, which has some sparse observations in \element[][12]CO, HCN and other molecules from ARO, is  shown as a red circle. Orange contours highlight the "cocoon" structure in \ion{H}{i} (contours at 15, 17 19 and 21 K km $\mathrm{s}^{-1}$, integrated between 2 and 4 km s$^{-1}$). The white compass indicates the  orientation in Galactic coordinates, so the Galactic plane is beyond the left side of the figure. Visible image showing dust reflection courtesy of Maurizio Cabibbo.}
         \label{fig:summary_obs}
\end{figure*}

We used observations from different telescopes combined with archival data to characterize  MBM 40.  Table \ref{tab:sumobs}  summarizes the observations.  Fig. \ref{fig:summary_obs} displays a $^{12}$CO map of the denser molecular structure of the cloud and both regions and single positions observed in CO, CH, and other species. Most of the observations were concentrated in four distinct regions, listed in Table \ref{tab:maps_position} with the same notation of Fig. \ref{fig:summary_obs} that we use in all of the following descriptions of our data set.

\begin{table*}[h!]
\caption{Summary of molecular spectroscopic observations, excluding region Z.}             
\label{tab:sumobs}      
\centering          
\begin{tabular}{c c c c c c l}
\hline\hline       
Molecule & $\nu$ (GHz) & Telescope & Location(s) & $\Delta$v (km s$^{-1}$) & beamsize & type\\
\hline  
\element[][12]CO & 115.271 & FCRAO & whole cloud & 0.05 & 47" & half beam-spacing map \\
\vspace{0.3cm}
 & & OSO & MAP1, 2, 3, 4 & 0.012 & 33" & full beam-spacing 9$\times$9 maps \\
\element[][13]CO & 110.201 & OSO   & MAP1, 2, 3, 4 & 0.013 & 34" & full beam-spacing 9$\times$9 maps \\
\vspace{0.3cm}
 & & ARO & MAP2 & 0.033 & 57" & half beam-spacing 3$\times$3 map \\
\vspace{0.3cm}
HCO$^+$ & 89.188 & ARO & MAP2, 4 & 0.041 & 71" & half beam-spacing 3$\times$3 maps \\
\vspace{0.3cm}
H$_2$CO & 4.830 & Arecibo & MAP1, 2 & 0.047 & 56" $\times$ 65" & full beam-spacing 5$\times$5 maps \\
\vspace{0.3cm}
HCN & 88.632 & ARO & MAP2 & 0.041 & 71" & single pointing (center of MAP2) \\
\vspace{0.3cm}
CS & 97.981 & ARO & MAP2 & 0.037 & 64" & half beam-spacing 3$\times$3 map \\
CH & 3.335 & Arecibo & whole cloud & 0.069 & 78" $\times$ 96" & full beam-spacing map \\ 
\hline\hline                  
\end{tabular}
\end{table*}

\begin{table}[h!]
\caption{Map centers of OSO observations. Each map is about 0\fdg07 wide.}             
\label{tab:maps_position}      
\centering                          
\begin{tabular}{c c c}        
\hline\hline                 
name & $\alpha_0$ (J2000) & $\delta_0$ (J2000) \\    
\hline                        
   MAP1 & 242\fdg7500 & +21\fdg7500 \\      
   MAP2 & 242\fdg6000 & +21\fdg9167 \\
   MAP3 & 242\fdg5500 & +22\fdg3000 \\
   MAP4 & 242\fdg5250 & +22\fdg0500 \\
\hline
\end{tabular}
\end{table}

\subsection{Onsala Space Observatory (OSO)}
\label{subsec:oso}
The Onsala Space Observatory (OSO), operated by Chalmers University of Technology in Göteborg, provides a 20 meter-class radome-enclosed antenna with millimeter and submillimeter capabilities (\citet{2015A&A...580A..29B}). The telescope is equipped with a 3 mm SIS mixer receiver (85 - 116 GHz frequency range, dual polarization). As backend, we used the OSA FFT spectrometer with the maximum resolution, 4.8 kHz for 2 $\times$ 156 MHz total bandwidths (one for each polarization) corresponding to $\Delta$v $= 0.012$ km s$^{-1}$ at 115 GHz (\element[][12]CO first rotational transition). The telescope has three different observing modes, position, beam and frequency switching. Because of baseline stability  considerations and target size, we used frequency switching mode with a frequency throw $\Delta \nu = 4$ MHz. The mean antenna temperature, T$_A$, for \element[][12]CO was about 10 K and for \element[][13]CO was about 3.5 K and hereafter we will use T$_A$ in the  OSO spectra. Dividing the antenna temperature by the antenna beam efficiency ($\eta_B \simeq 0.3$) we obtain the main beam temperature. In spring 2016 we made four different maps (numbered 1-4 in Fig. \ref{fig:summary_obs}), both in \element[][12]CO and \element[][13]CO, each with 81 pointings, arranged in a 9$\times$9 pattern made by full beam sampling.

\subsubsection{OSO CO data}
Our new CO spectra from OSO were obtained in March 2016.  The observations were performed in frequency-switching mode (FSW, i.e. the Local Oscillator (LO) changes the central frequency five times per second). In FSW mode an OFF position is not observed thereby doubling the on-source time. However, this procedure creates two copies of the same line in each spectrum so a folding procedure is mandatory. For each position observed, we averaged the spectra and then folded the data. We then selected only 400 channels centered on the cloud line, eliminating the spurious negative antenna temperature  lines created by the folding procedure and, for \element[][12]CO, the telluric mesospheric line (see, \cite{2022ApJS..262....5D}). Because \element[][12]CO and \element[][13]CO have slightly different velocity resolution, we interpolated each spectra using an even-spaced velocity array with $\Delta$v = 0.02 km s$^{-1}$. The final velocity resolution of \element[][13]CO is a little worse than the original data and for \element[][12]CO is slightly improved. However, the difference between uninterpolated and interpolated data is negligible for our study. Typical rms noise level for the reduced \element[][12]CO was $\sim$ 1 K and for \element[][13]CO was $\sim$ 0.7 K.

\subsection{Arecibo Radio Telescope}\label{subsec:arecibo}
The William E. Gordon Radio Telescope, located near Arecibo, Puerto Rico, provided H$_2$CO (4.830 GHz - Arecibo project A3261) and CH (3.335 GHz - Arecibo projects A1659 and A1708) archival data of MBM 40. The 305 meter primary reflector diameter allows good spatial resolution even at low frequency, showing a beam size of $\sim$ 1' for H$_2$CO and slightly larger (1.3' $\times$ 1.6') for CH, making these observations comparable with the high frequency maps. We used the S-high and C band, dual linear polarization receivers with bandwidth 0.78125 MHz for the CH observations and 1.5625 MHz for H$_2$CO. The velocity resolutions are 0.068 km s$^{-1}$ at 3.3 GHz and 0.047 km s$^{-1}$ at 4.8 GHz. We covered MAP1 and MAP2 in H$_2$CO, using for each 25 pointings arranged in 5$\times$5 pattern. The coverage of the H$_2$CO maps is the same as the OSO map, with a slightly lower spatial resolution. CH spectra come from older observations (\cite{2010AJ....139..267C}) over most of the denser parts of the cloud. Though the CH observations were done for another project and are not arranged in any particular pattern inside each of the 4 regions, each map is covered by 7 - 10 CH pointings.  In addition, we reanalyzed a diagonal strip (shown as blue diamonds in \ref{fig:summary_obs}) as described below.

We also used the GALFA-HI (Galactic Arecibo L-band Feed Array HI) narrow band data archive for \ion{H}{i} (see \cite{2011ApJS..194...20P}; \cite{2018ApJS..234....2P}), an extended survey between $-1 \degr \lesssim \delta \lesssim 38 \degr$ with angular resolution of $4\arcmin$ and $0.184 \ \mathrm{km \ s}^{-1}$ spectral resolution. Each spectrum has 2048 channels in the velocity range of $\left|\mathrm{v_{LSR}}\right| \lesssim 188$ km s$^{-1}$; the \ion{H}{i} emission in the environs of MBM 40 is located between -4 to 6 km s$^{-1}$.

\subsubsection{H$_2$CO 4.56 GHz data}
The new H$_2$CO data were obtained in 2018 July-August using the 305-m reflector at the Arecibo Observatory. An H$_2$CO 4.8 GHz map of the cloud at 6' resolution was presented by \cite{1996ApJ...465..825M}. The C band dual linear polarization receiver was used with typical system temperatures of 25-30 K on the sky. The beam size at 4.8 GHz was 56" $\times$ 65". The backend was the Wideband Arecibo Pulsar Processor (WAPP)  configured into eight sections each with a bandwidth of 1.5625 MHz over 2048 channels.  Two of the WAPP sections were centered on the 4830 MHz H$_2$CO 1(1,0)-1(1,1) transition which consists of six hyperfine components from 4829.6412 to 4829.6710 MHz. The remaining six sections were centered on hyperfine transitions of OH, HCOOH, and CH$_2$CNH ranging from 4.8 - 5.0 GHz, all of which resulted in non-detections. The velocity resolution of the H$_2$CO data was 0.047 km s$^{-1}$ and maps were made in a 5$\times$5 pattern covering the region of the 9$\times$9 OSO CO observations.  Because of time constraints, only MAP1 and MAP2 were completed in H$_2$CO.  The observations were made observing  on-source only, with the WAPP frequency band structure removed by subtracting a first order polynomial. Typical 1$\sigma$ rms noise levels for the reduced spectra were between the range 8-17 mK. The results are given in units of antenna temperature, T$_A$. Conversion to the brightness temperature, T$_B$ is done by dividing by the beam efficiency which was measured to be $\sim$0.6.

\subsection{Arizona Radio Observatory (ARO)}\label{subsec:aro}
Several other molecules were observed using the 12 meter millimeter-wave radio telescope of the Arizona Radio Observatory (ARO) on Kitt Peak during 2016 April-May. Observations were performed using the 3 mm ALMA Type Band 3 dual-polarization receiver combined with Millimeter Autocorrelator (MAC) spectrometer and 250 kHz filter banks, the latter used as backup for the MAC. The MAC was configured to have 200 MHz total bandwidth (150 MHz of usable bandwidth) over 8192 channels for a frequency resolution of 24.4 kHz. The velocity resolution and beamsize of each transition are shown in Table \ref{tab:sumobs}. The antenna temperature scale, T$_A^*$ (see \cite{1981ApJ...250..341K}), is set by the chopper wheel method and is given at the telescope as T$_R^*$, the antenna temperature corrected for spillover and scattering. Conversion to the main beam brightness temperature, T$_{mb}$, is via T$_R^* / \eta_{mb}$, where $\eta_{mb}$ is the main beam efficiency (at 115 GHz, $\eta_{mb} \approx 0.85)$. Because MBM 40 fills the beam, we assume $\sim$1 for the beam filling factor.

The new data from ARO comprise CO, HCO$^+$, C$_2$H, C$_3$H$_2$, HCN, CS observed in MAP1, MAP2, MAP4 and MAPZ. Because we have CO spectra with higher frequency resolution from OSO, we used the ARO spectra only for MAPZ, the region we had not observed with OSO. All data were taken using frequency switching.  The integrations used half-beamwidth spacing maps (for \element[][12]{CO}{}), and deep integration and 5-point cross (for other species). Each spectrum was folded and baseline corrected using first or second order polynomial. The gallery of detected emission is displayed in Fig. \ref{fig:summary}.

\subsection{Five College Radio Astronomy Observatory (FCRAO)}\label{subsec:fcrao}
Although the principal CO data used in this study is from OSO, outside the four numbered regions we used earlier CO observations from the Five College Radio Astronomy Observatory (FCRAO) data, analyzed and discussed by \cite{2003ApJ...593..413S}. The observations  of \element[][12]CO (1-0) were obtained with on-the-fly SEQUOIA mapping and frequency switching. The datacube is composed of over $24 \, 000$ spectra with a velocity resolution of 0.05 km s$^{-1}$, and an average rms noise of 0.7 K. The spectra are corrected for scattering and spillover. We assumed a beam filling factor $\sim$1. For a detailed  discussion of FCRAO observations see \cite{2003ApJ...593..413S}.

\subsection{Infrared imaging archival data from {\it WISE} and {\it Planck}}
To study the cloud's dust component we used archival data from {\it Planck} and {\it Wide-field Infrared Survey Explorer (WISE)}. {\it Planck} employed a variety of infrared, millimeter and sub-millimeter detectors to characterize the spectral distribution of the Galactic foreground emission to remove it  from the Cosmic Microwave Background (see \cite{2020A&A...641A...3P} and \cite{2020A&A...641A...4P}). We use the 545 GHz channel, which traces cold thermal dust.  {\it WISE} (see \cite{2010AJ....140.1868W}) mapped the whole sky at 3.4, 4.6, 12 and 22 $\mu$m with improved sensitivity near the ecliptic poles. The high sensitivity, coupled with a good angular resolution, allows a detailed analysis of MBM 40 and the outlying gas in these wavelengths.

\section{Gas tracers and dust}

\subsection{Tracing diffuse gas: CH and HCO$^+$}
\begin{figure}
  \centering
  \includegraphics[width=\hsize]{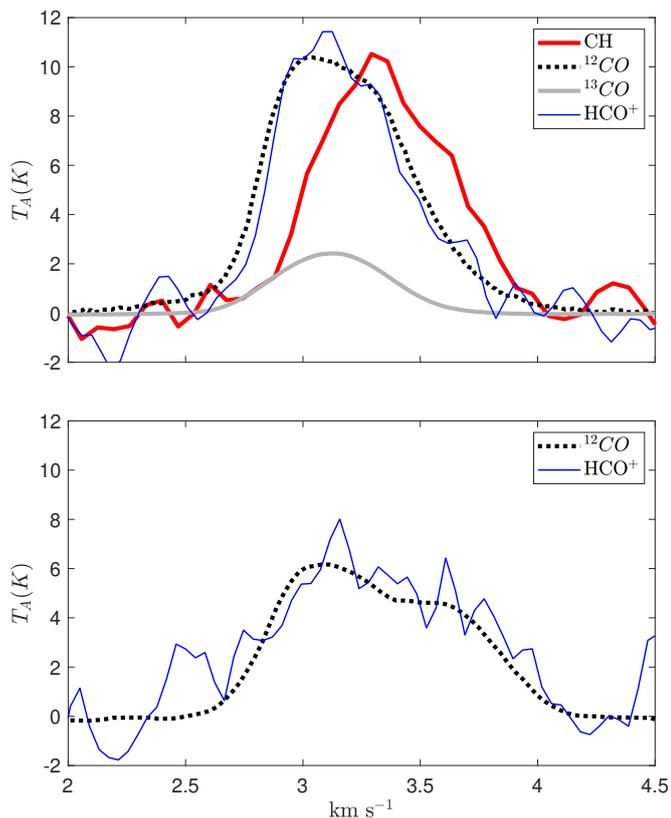}
     \caption{\textit{(Upper panel.)} MAP2 molecular lines. To increase the S/N, all spectra of the same species were averaged. The \element[][13]{CO}{} traces predominantly denser gas, so present a narrower linewidth respect to diffuse gas tracers. \textit{(Lower panel.)} MAPZ molecular lines. There is only one CH spectrum near MAPZ (position 11 in Fig. \ref{fig:summary_obs}) but it shows a very low S/N so we did not plot it here. All line intensities are normalized to \element[][12]CO peak antenna temperature, except for \element[][13]{CO}{}, which is shown without any rescaling.}
        \label{fig:ch_hcoplus}
\end{figure}

\element[][12]{CO}{} is widely used as a tracer of molecular hydrogen due to its ubiquity, although the J=1-0 line is often optically thick and the relation between CO antenna temperature and N(H$_2$) changes even within the same molecular cloud (\cite{1998ApJ...504..290M}, \cite{2015ApJ...811..118R}). In contrast, CH shows a linear relation between N(CH) and N(H$_2$) (\cite{1976ApJS...31..333R}, \cite{1986A&A...160..157M}, \cite{2002A&A...391..693L}), both in diffuse and translucent clouds where A$_V < 5$ and is an excellent tracer for low density gas ($n < 10^3$ cm$^{-3}$) at cloud boundaries. Because the peak temperature of CH in diffuse clouds is typically a few tens of mK, observations require long integration times. The upper panel of Fig. \ref{fig:ch_hcoplus} shows the \element[][12]CO, \element[][13]{CO}{}, HCO$^+$ and CH observations in MAP2, where we have the full set of tracers. \element[][12]{CO}{}, HCO$^+$ and CH have compatible linewidths, broader than \element[][13]{CO}{} and CS lines. \element[][13]{CO}{} is tracing predominantly the denser gas, while CH traces mainly the outer, diffuse gas. \element[][12]{CO}{} seems to trace both diffuse and dense gas, presenting a wide profile with the right wing very similar to the right wing of CH. HCO$^+$ is a precursor of \element[][12]{CO}{}, so the profile is nearly identical to \element[][12]{CO}{}. The peak velocity shift between \element[][12]{CO}{} and CH is discussed in subsection \ref{subsec:shear}.
In other maps and positions (see Fig. \ref{fig:summary}) we don't have a complete set of molecules, however the HCO$^+$, CH and \element[][12]{CO}{} show the same width. In MAP4, where the \element[][12]{CO}{} evidently shows a double peak, the CH profile is slightly shifted to the left, compatible with the \element[][12]{CO}{} blueshifted wing at 2.5 km s$^{-1}$; also HCO$^+$ shows a broadened profile towards lower velocities. In MAPZ, the \element[][12]{CO}{} traces some gas at $\sim$3.8 km s$^{-1}$, picked up also by HCO$^+$ (lower panel of Fig. \ref{fig:ch_hcoplus}). Consequently, we have complementary molecular tracers for the different density regimes of the cloud.

\subsection{Dust - gas relation}
\begin{figure}[h!]
  \centering
  \includegraphics[width=1\hsize]{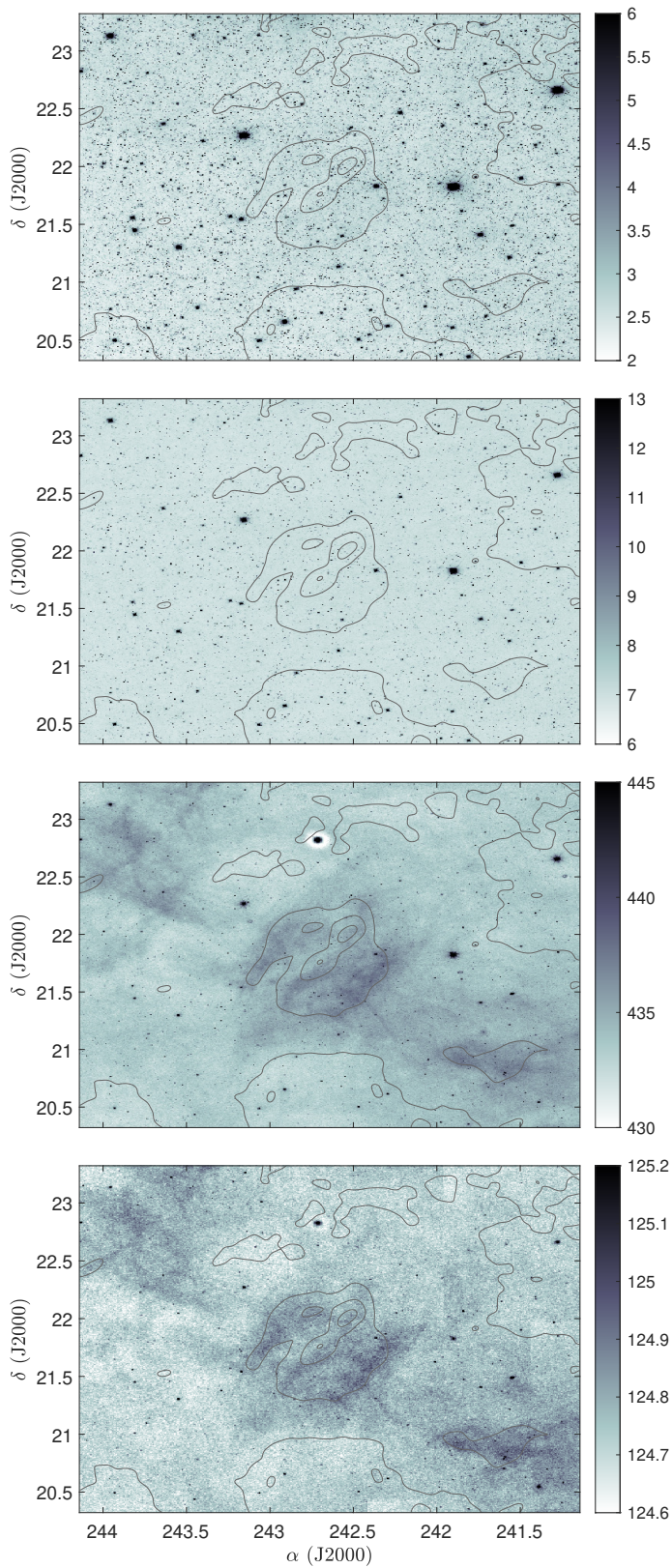}
     \caption{{\it WISE} images of MBM 40: From the top: 3.4, 4.6, 12 and 22 $\mu$m. Over each image contours from 545 GHz {\it Planck} channel which traces cold dust. Each {\it WISE} image is histogram equalized to most enhance the details (uncalibrated; colorbar units in DN). Contours spanning from 1 to 4 MJy sr$^{-1}$.}
        \label{fig:WISEPlanck}
\end{figure}

Fig. \ref{fig:WISEPlanck} displays the four {\it WISE} channels with superposed {\it Planck} 545 GHz contours. The shorter  wavelength channels (3.4 and 4.6 $\mu$m) show no emission from the cloud, whereas the 12 and 22 $\mu$m channels show diffuse emission that corresponds to only the peripheral gas.  The 12 $\mu$m channel shows an anti-correlation between the {\it WISE} and {\it Planck} emission: the bulk of the molecular cloud, clearly seen by dust ({\it Planck}) shows virtually no emission in 12 $\mu$m, and where the 545 GHz emission is lower (i.e. in the south-west part) the 12$\mu$m emission is higher. There are also some areas visible at 12 $\mu$m but not in the  545 GHz image.

The 12 $\mu$m channel traces mainly PAHs emission, excited by the ISRF (Interstellar Radiation Field); the 22 $\mu$m channel traces warm dust ($\sim$ 140 K) which is heated by diffuse radiation. The MBM 40 hairpin structure is not visible in these two images, suggesting that it is optically thick to the UV and shields the inner gas. The lack of emission at 22 $\mu$m can be  explained by the low temperature of the dust (about 18 K), as seen in the dust temperature map from {\it Planck} (\cite{2022A&A...668L...9M}).  Near MAP1 there is a strong correlation between \ion{H}{i} and the PAH emission, as they are both subdominant relative to the molecular gas.

\begin{figure*}
  \centering
  \includegraphics[scale=0.4,bb=0 0 1300 450,clip]{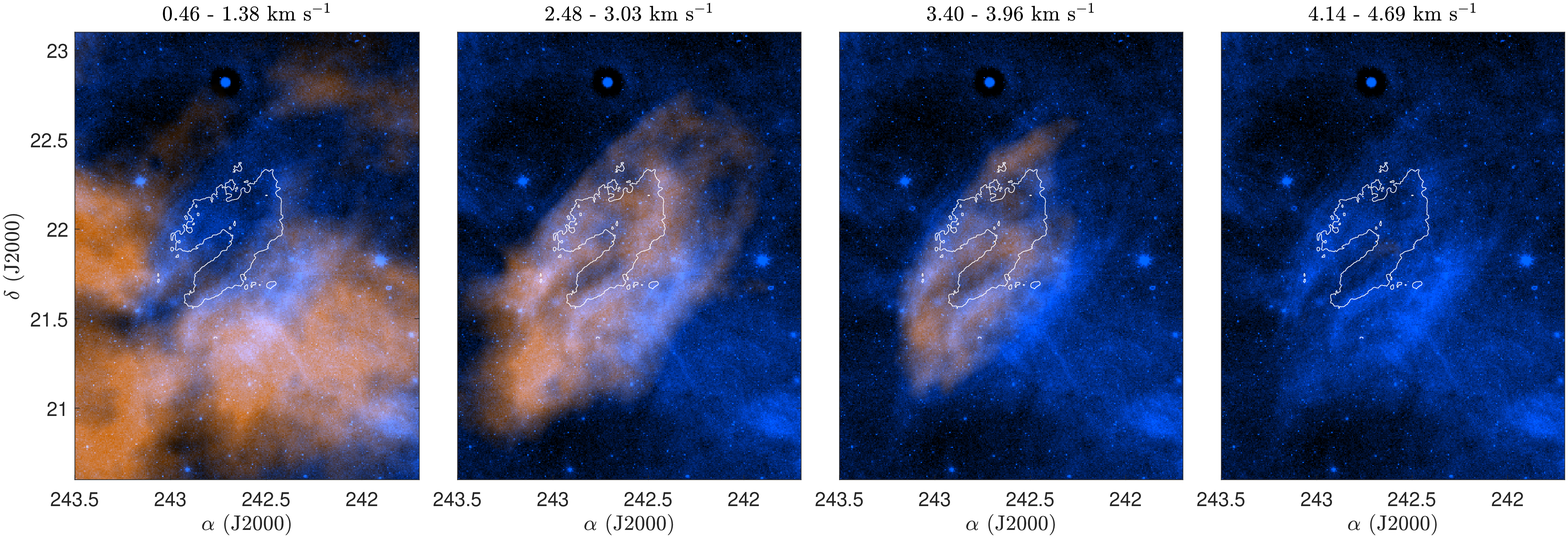}
     \caption{Set of composite images from GALFA-HI and {\it WISE} 12 $\mu$m. The {\it WISE} image is the same in each panel, and the brightness temperature averaged over the given velocity range is shown in dark blue. The title over each panel indicates the velocity range over which the GALFA-HI datacube has been integrated, and is shown in orange. Both {\it WISE} and GALFA-HI monochromatic colormap are the same in all panels and track the intensity. {\it WISE} pixel values are in Digital Numbers (DNs) without any calibration, so we selected colormap limits to enhance the cloud structure. GALFA-HI colormap limits are 20 and 32 K. White lines show the \element[][12]{CO}{} integrated intensity contour 1 K km/s.}
        \label{fig:WISEhi}
\end{figure*}

The band-integrated emission mapped by {\it WISE} has no velocity information, but the individual \ion{H}{i} channels permit a dynamical separation of different regions and density regimes (see Fig. \ref{fig:WISEhi}). 
The first panel, for the velocity interval (0.46 - 1.38 km s$^{-1}$) shows the gas in the outer region of MBM 40 indicating that the surrounding dust and PAH emission are not directly linked with the densest portions of the cloud.
In this interval the \ion{H}{i} also picks up the gas along the bar that crosses the southern part of the molecular cloud and foreground/background gas around $\alpha =$ 243\fdg3, $\delta =$ +21\fdg1. Between 2.5 and 3.5 km s$^{-1}$ \ion{H}{i} traces only the cocoon structure of the cloud (for a discussion of MBM 40 atomic gas structure see \cite{2003ApJ...593..413S}). The western ridge shows a more jagged shape than the eastern ridge, which displays a steeper gradient  in both velocity intervals.  We discuss this filamentary structure of the ambient gas in Appendix \ref{app:periph_struct}.

\subsection{H$_2$CO}
We covered MAP1 and MAP2 using Arecibo with 5$\times$5 footprints (see Appendix \ref{app:h2co}). H$_2$CO 4.83 GHz (1$_\mathrm{11}$ - 1$_\mathrm{10}$) is intrinsically wider than other molecular lines used in this work, because of its six hyperfine lines. The line is seen in absorption at densities below about $10^6$ cm$^{-3}$  and goes into  emission for higher densities.  Thus a weak absorption might be indicative of a higher density. Here, however, weak absorption is due to low density because MBM 40 is a diffuse cloud. The line is a good tracer of low density core regions but it does not pick up the gas from the outer boundaries unless integrations of several hours per point are attempted (see \cite{1993ApJ...402..226M}). This conclusion is reinforced by the spectra in MAP1 and MAP2, where the H$_2$CO fades out at the map boundaries, but  \element[][12]CO emission is still present. The H$_2$CO raw spectra are all in absorption, but in this work we  show them as emission  to ease the comparison with other tracers.

Figs. \ref{fig:map1_h2co_co} and \ref{fig:map2_h2co_co} show that the peak velocity of H$_2$CO main line is shifted from the \element[][12]CO central velocity. A plausible explanation is that \element[][12]CO line is a blend of two different gas regimes, one very diffuse at the boundary and one denser that follows the central spine of the western flow.  In  this scenario the H$_2$CO traces only the denser gas, given our relatively short integration times.

\section{Data analysis: Dynamics}\label{sec:data_analysis_dyn}
In this section we focus on dynamics inside MBM 40, primarily analyzing the turbulent field and its implications.

\subsection{H$_2$CO dynamics}
Because the 4830 MHz formaldehyde line is actually formed by six hyperfine different components, the profile analysis is more complicated because a satellite hyperfine line might easily mimic a secondary velocity component. Consequently, it is crucial to model the theoretical profile to separate real dynamic signatures from the intrinsic ensemble shape.

If we assume that the H$_2$CO and \element[][12]{CO}{} are well mixed, the \element[][12]{CO}{} line profiles -- interpreted as a probability distribution of the velocity fluctuations (see \ref{sub:single_pdf}) -- provide convolution kernels that can be used with theoretical profiles of H$_2$CO, based on the laboratory intensities of the hyperfine components, to match the observations.

Fig. \ref{fig:summary} shows that the molecular tracers may be divided in two groups: \element[][12]{CO}{}, CH, HCO$^+$ show generally broad lines, while \element[][13]{CO}{} and CS show narrow lines. The formaldehyde is the outlier. In this section we apply a theoretical procedure to understand the gas regime mapped by H$_2$CO.

As a first step we constructed a purely thermal broadened theoretical profile of H$_2$CO. We assume LTE and the same intrinsic thermal width for each hyperfine component. In order to demonstrate the implications of this assumption, we performed some simulations with different H$_2$CO column densities using the RADEX code (\cite{2007A&A...468..627V}), confirming that the line is optically thin below N(H$_2$CO) $\sim$ 10$^{14}$ cm$^{-2}$, obviously well above our limit.

\begin{figure}
  \centering
  \includegraphics[width=\hsize]{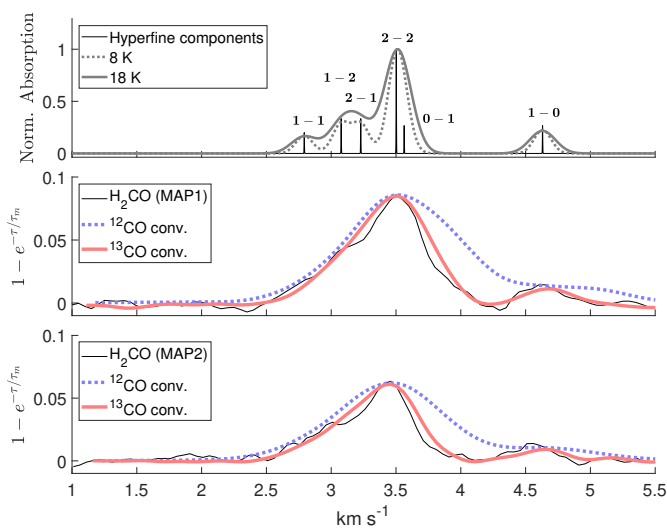}
     \caption{H$_2$CO profile modeling, displayed as emission. The top panel shows the theoretical relative intensities of the hyperfine components without any thermal broadening (thin black line) with F $\rightarrow$ F' hyperfine transition indicated. Two different thermal broadened profile are also shown. In the central and lower panel are shown the observed H$_2$CO profile (black thin line) with theoretical profiles convolved by the \element[][12]{CO}{} and \element[][13]{CO}{} -based kernel.}
        \label{fig:conv_h2co}
\end{figure}

The theoretical H$_2$CO optical depth is assumed to be given by a sum of Gaussians each weighted with the laboratory oscillator strength of the related hyperfine transition,
\begin{equation}
    \tau = \sum_n \tau_{(0,n)} \mathrm{exp} \left[- \left(\frac{\mathrm{v-v}_{0,n}}{2\sigma}\right)^2 \right] \ \ \ \ ,
    \label{eq:tau}
\end{equation}
\noindent
where $\tau_{(0,n)}$ is the relative optical depth of the $n$-th hyperfine line (see \cite{1970ApJ...161L.153T} and the table therein),
$\mathrm{v}_{0,n}$ is the relative velocity displacement of each component relative to the main line, and $\sigma$ is the width of the line. The upper panel of Fig. \ref{fig:conv_h2co} shows the hyperfine lines ensemble of formaldehyde at 4830 MHz. The thin black line is the relative intensities of the lines normalized to the main line, that is the spectrum without any thermal broadening, while the dotted and bold thick lines show the resulting LTE spectrum with thermal broadening corresponding to 8 and 18 K, respectively. A cloud temperature of 18 K (see \cite{2022A&A...668L...9M}), corresponding to a sigma $\sim$ 0.07 km s$^{-1}$, under LTE conditions cannot explain the actual H$_2$CO profile. A temperature of 40 K would be necessary to fit the observed profile, but this temperature is incompatible with that derived from \textit{Planck} data. To model the dynamical broadening, we use both \element[][12]{CO}{} and \element[][13]{CO}{} lines to indicate the velocity distribution of the gas. Then the theoretical H$_2$CO profile is given by

\begin{equation}
    \phi(\mathrm{v})_{\mathrm{theo}} \propto \mathrm{exp}\left[-\Phi\ast \left(\frac{\tau}{\tau_{m}}\right) \right] \ \ \ \ ,
    \label{eq:theo_prof}
\end{equation}
\noindent where $\Phi$ is the normalized profile from the $^{12}$CO or $^{13}$CO at the same position assuming it is the PDF of the line of sight turbulent motions within the beam that is convolved with $\tau/\tau_m$, where $\tau_m$ is a scaling factor for the optical depth.

The central and bottom panels of Fig. \ref{fig:conv_h2co} show the theoretical H$_2$CO profiles for MAP1 and MAP2 as black solid lines, and the observed profiles are shown by the solid red lines, the dashed black line is the theoretical profile convolved with the \element[][12]CO-based kernel and the solid blue line is with the \element[][13]CO-based kernel, both calculated using the same H$_2$CO theoretical profile (computed by Eqs. \ref{eq:tau} and \ref{eq:theo_prof}).

Because \element[][12]CO and \element[][13]CO have no  hyperfine structure, their line profiles can be treated as the probability distribution function of the gas velocity (see subsection \ref{sub:single_pdf} for a detailed discussion). The \element[][13]{CO}{} is optically thin in diffuse interstellar environments, so the line profiles are sampling the gas velocity distribution along each line of sight. The \element[][12]{CO}{} is much more abundant than \element[][13]{CO}{} and quickly becomes optically thick, as along the hairpin in MBM 40. However, the ratio between \element[][13]{CO}{} and \element[][12]{CO}{} shows that \element[][12]{CO}{} picks up both optically thick gas (center of the line and the blueshifted wing) and optically thin gas (the redshifted wing): the departure between H$_2$CO profile and the theoretical profile convolved with the \element[][12]{CO}{}-based kernel is seen around 4 km s$^{-1}$, where \element[][12]{CO}{} is tracing the optically thin gas. 

The formaldehyde profile convolved with the \element[][13]{CO}{}-based kernel is compatible with the observed H$_2$CO line, while  the \element[][12]{CO}{}-convolved profile is not, indicating that the formaldehyde is tracing the same denser gas component as \element[][13]CO. The synthetic profile is also in good agreement in the non-Gaussian wings. In contrast, the \element[][12]CO convolved profile is too wide compared with formaldehyde, indicating that \element[][12]CO is formed over a broader velocity range. With this analysis, and as shown in Fig. \ref{fig:summary} we find that the \element[][12]CO, CH, HCO$^+$ transitions are tracing the same gas dynamics, whereas \element[][13]CO, H$_2$CO and CS are tracing the denser component.

\subsection{Shear flow}\label{subsec:shear}
Although MBM 40 has no internal star formation, the spectra show superthermal line widths that  indicate  turbulent motion. However, as shown by \cite{1998PhRvL..80.2754M}, turbulence decays on a short timescale, so a continuous source of kinetic energy is necessary. Without star formation, this source must be external.  Large scale shear flows are likely the best candidates (see \cite{1999ApJ...512..761L}). In MBM 40 we found evidence for a large shear flow that runs through western filament, visible in \element[][12]CO and in CH. This shear flow is also visible in \ion{H}{i}, but is partially masked  by diffuse external gas that blends with the peak velocities.

\begin{figure}
  \centering
  \includegraphics[width=\hsize]{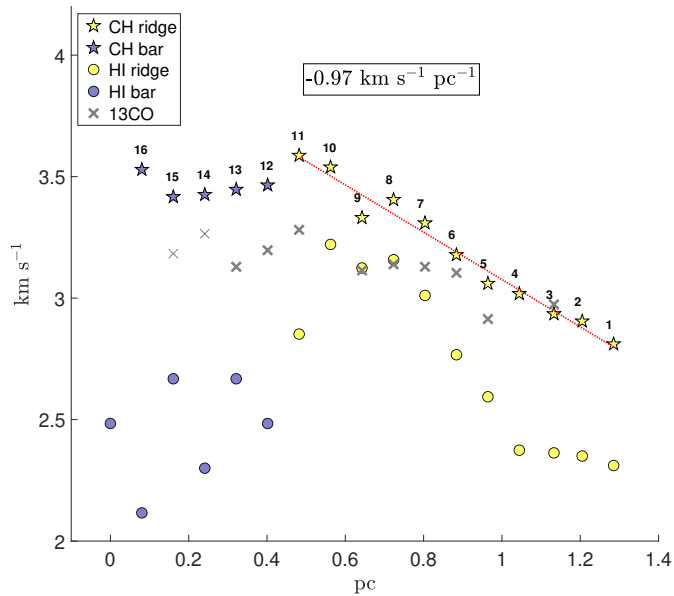}
     \caption{CH, \ion{H}{i}, \element[][13]CO Gaussian-fitted velocity along the west ridge of the cloud. Each position is quoted as in Fig. \ref{fig:summary_obs}, blue diamonds. The first two \element[][13]CO peak velocities are shown in thin line because here the S/N is too low to trust the Gaussian fit. We converted angular distance into physical distance using the cloud distance of 93 pc (see \cite{zucker2019}). In CH, a linear correlation was found (red dotted line).}
        \label{fig:diagonal_ch}
\end{figure}

Fig. \ref{fig:diagonal_ch} displays a summary of shear analysis using \ion{H}{i}, CH and \element[][13]CO. For each CH and \element[][13]CO spectrum, we performed a single Gaussian fit and plotted the peak velocities versus distance along the western ridge. The starting point is below the 16th point in Fig. \ref{fig:summary_obs}, where the \ion{H}{i} velocity component associated with the cloud appears. Because \ion{H}{i} generally shows more complex profiles, we performed a Gaussian decomposition as described in our previous work (\cite{2022A&A...668L...9M}) and plotted just the peak velocity of the component associated with MBM 40. The shear is  visible in CH and \ion{H}{i}, while \element[][13]CO shows a poorer correlation. The velocity shear exhibits an abrupt change at positions 11-12 in the lower part of the cloud.  This may be a superposition of foreground and background gas associated with a bar-like structure that crosses MBM 40, distinctly visible in the extended \ion{H}{i} GALFA-HI map and  in the visible image (see Fig. \ref{fig:summary_obs}). Unfortunately, no \element[][12]CO data are available for this bar.\footnote{We examined the Harvard-Smithsonian CfA millimeter-wave CO survey  \cite{2022ApJS..262....5D}. Although the spatial sampling is coarse,  there is  no detected \element[][12]CO emission along the bar.}

\begin{figure}
  \centering
  \includegraphics[width=1.0\hsize]{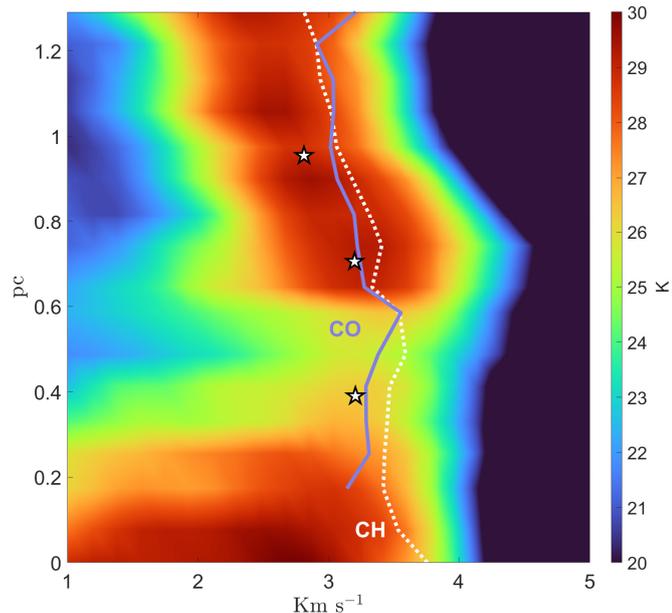}
     \caption{Shear flow in the western ridge of MBM 40. The pseudocolor image is \ion{H}{i}. \element[][12]CO peak velocities are plotted in blue and CH in dotted white. Stars indicate the mean \element[][13]CO peak velocity observed with Onsala telescope in MAP1, MAP2 and MAP4. \element[][12]CO missing points below 0.2 pc shows too low S/N.}
         \label{fig:shear}
\end{figure}

The southernmost CH spectra show a double peak emission, supporting the idea that the line of sight intercepts two velocity components, one of which arises from the bar seen in IR images. We therefore apply a double Gaussian decomposition for each CH profile, and then consider only the redshifted peak velocity, which is associated with the MBM 40 western ridge. The result is displayed in Fig. \ref{fig:shear}.
Using only the redshifted components for CH and \element[][12]CO, the shear is more evident without any abrupt velocity changes in the lower part of the cloud.  It systematically follows  the spine of the ridge traced by \ion{H}{i}. We note that in the southernmost part of the cloud, the \ion{H}{i} is picking the gas from the bar, which shifts the central velocity toward 2 km s$^{-1}$.

The systematic offsets between the CO and CH velocity centroids along the spine are the same as described in \cite{2010AJ....139..267C}, who explained it as arising from two different sampled density regimes.  In that case, the difference along the spine appears to indicate a transverse velocity gradient that may be a genuine shear (that is, not from topology, see subsection \ref{subsec:long_shear}) of about 1 km s$^{-1}$pc$^{-1}$. If this is dynamical, we suggest that this systematic differential velocity represents  motion induced by driving by the external neutral atomic gas.

The neutral atomic hydrogen shows a complex velocity distribution.  The emission at v$_{LSR}$ < 2 km s$^{-1}$ is from diffuse gas that is from an extended structure that continues toward the east of the cloud. The central spine of the MBM 40 western ridge is visible between 2 and 4 km s$^{-1}$, showing a shift toward higher positive velocities -- from 2.6 km s$^{-1}$ to 3.3 km s$^{-1}$ -- on a length scale of 0.6 pc.

\element[][13]CO was observed only in MAP1, 2 and 3, so no separate shear analysis can be performed. However, in Fig. \ref{fig:shear}, we show the mean peak velocity for each map (gray stars): \element[][13]CO traces only the densest molecular material embedded within atomic gas, showing a departure from CH and \element[][12]CO above 0.9 pc.

\begin{figure*}
  \centering
  \includegraphics[scale=0.43,bb=0 0 1300 590,clip]{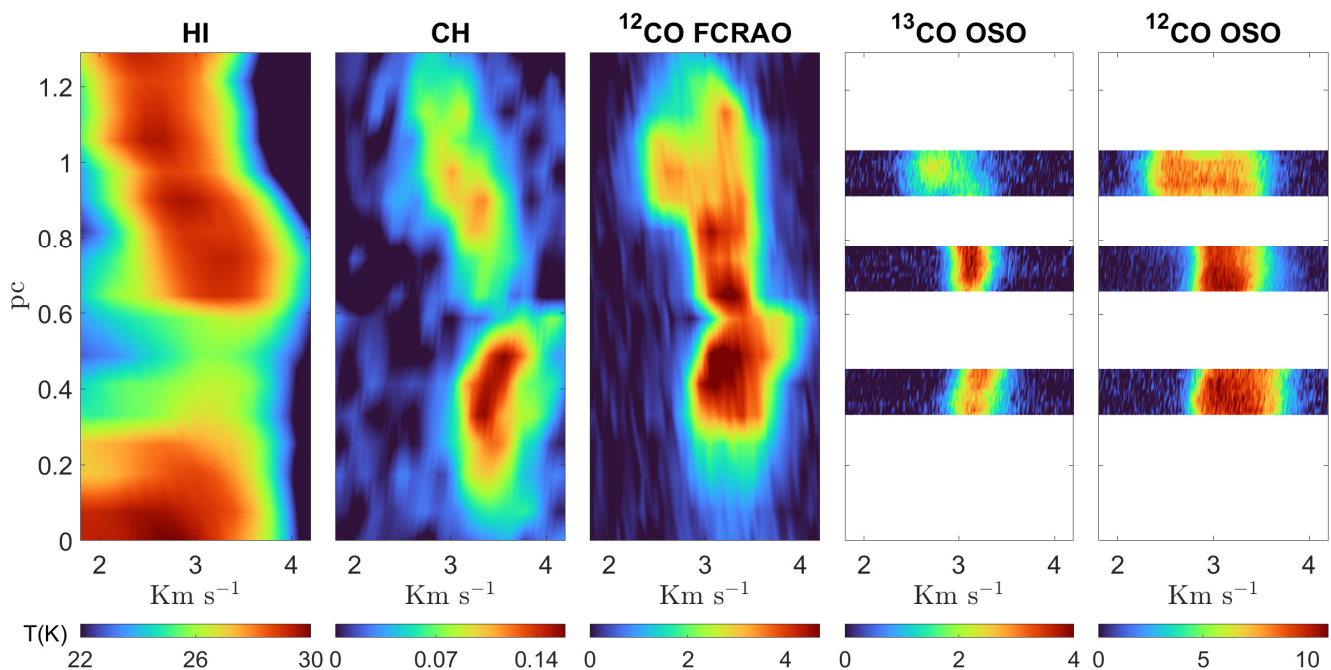}
     \caption{Shear flow traced by \ion{H}{i}, CH, \element[][12]CO and \element[][13]CO. Note the two different dynamical regimes (exchanging at 0.6 pc) picked by CH and \element[][12]CO observations, and the CH which traces better the less dense and fainter gas in the lower part of the cloud, where \element[][12]CO is virtually absent, at least at the W(CO) $\ge$ 1 K km s$^{-1}$ level.}
        \label{fig:three_shear}
\end{figure*}

Fig. \ref{fig:three_shear} presents the longitudinal P-V plots for \ion{H}{i}, CH, \element[][12]CO and the incomplete \element[][13]CO, without component decomposition. For each CH position we selected the nearest \ion{H}{i} and \element[][12]CO spectrum, and then created the P-V image.  These plots are generated selecting evenly-spaced spectra over the western ridge to map the shear, not  by collapsing spectra along the same right ascension. The peak velocity is not useful because the profiles are complicated due to  different dynamics and even superposition of nearly independent gas regimes (i.e. in atomic hydrogen spectra, where foreground and background gas is picked up).  Instead, we use the profile weighted centroid velocity (\cite{1985ApJ...295..466K}) for each point in the maps
\begin{equation}
   u_c = \int_{line} T_A(u)u  \ du \ \Big /\int_{line}T_A(u) \ du \ \ \ 
   \label{eq:cent}
\end{equation}
where $T_A$ is the antenna temperature in velocity channel $u$. The integration of the observed profile is performed between finite values where the line has reached noise level, removing  the left/right baselines.
Two different gas regimes are evident in all cases except  \element[][13]CO, with a break near 0.55 pc. Furthermore, in \element[][12]CO, above 0.9 pc, a bifurcation is present, clearly visible also in single spectra taken with Onsala telescope in MAP4, where in CH is not present. We return to this below.

\subsection{Probability distribution functions (PDF) of centroids}
As shown previously, the \element[][12]CO spectra show complex profiles that are significantly wider than thermal motion providing a strong indication of turbulent motion inside the cloud (see \cite{1990ApJ...359..344F} \cite{1999ApJ...524..895M}, \cite{2006A&A...457..197S}), and non-Gaussian wings that indicate intermittency, large deviations in velocity that are rare events occurring with a higher frequency than in an uncorrelated process (\cite{1990ApJ...359..344F}, \cite{1995A&A...293..840F}, \cite{1998A&A...331..669F}, \cite{2009A&A...507..355F}). They may be detected with the centroid Probability Distribution Function (PDF) and even in single profile (see the next subsection for discussion). 
 \cite{1984ApJ...286..255K} (and subsequent papers, \cite{1985ApJ...295..466K}, \cite{1985ApJ...295..479D}) introduced  a statistical approach to study interstellar turbulence dynamics, using the autocorrelation functions (ACFs) of centroids. 

The PDF of velocity centroid increments is a valuable tool for the study of gas kinematics and can be used to characterize the turbulence scale (see \cite{2000tufl.book.....P}). Ordinarily, the centroids are used without removing any large scale trends (see \cite{2002A&A...390..307O}, \cite{2003A&A...412..417P}) or with some detrending (\cite{1994ApJ...429..645M}, \cite{1999ApJ...512..761L}). If some ordered flow is present, such as in the west ridge of MBM 40, detrending is crucial because the large scale velocity correlation can mask that from turbulence. Even for small  maps, not removing any systematic shear can cause problems since the large scale flow  can became dominant precisely where the turbulence starts showing differences from Gaussian processes.  The PDF corrected for the CH-traced gradient is then computed as the histogram of

\begin{equation}
\delta \mathrm{v} (\mathbf{r},\delta \mathbf{r}) = u_c(\mathbf{r})-u_c(\mathbf{r}+\delta \mathbf{r})
\end{equation}
\noindent
where $\mathbf{r}$ is the position in  the map and $\delta \mathbf{r}$ is the lag  that is the shift of the subtracted map from the original one. In the case of a raster map, $\mathbf{r}$ and $\delta \mathbf{r}$ should be decomposed in the indexes over the map, $(i,j)$ and $(\Delta i, \Delta j)$. The PDF is formed from fewer points for large lags because of the finite size of the map.
The PDF is then a histogram of velocity differences as a function of the lags and, for turbulence, the PDF of the centroids will show non-Gaussian wings at small lags that  relax to a Gaussian with large lags in the limit of uncorrelated values.

\begin{figure}
  \centering
  \includegraphics[width=0.8\hsize]{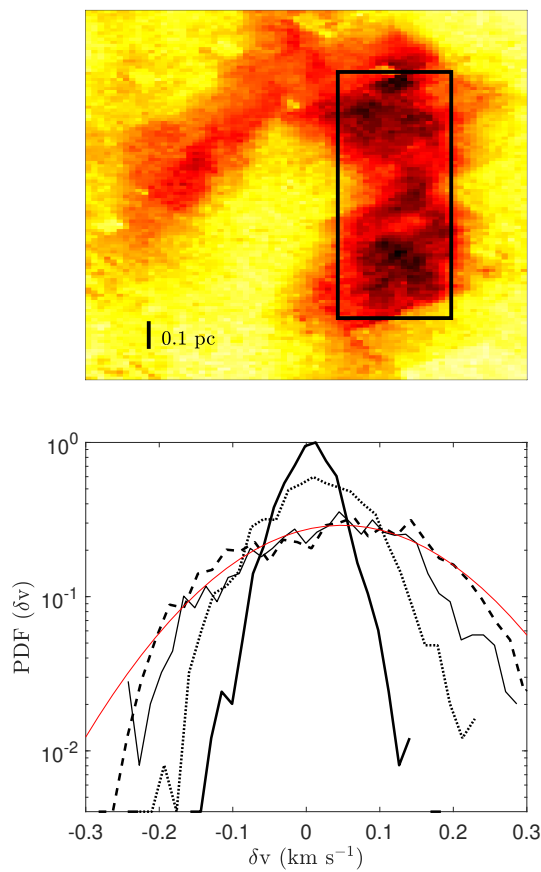}
     \caption{\textit{(Upper panel).} Selected region in west ridge from \element[][12]CO FCRAO data for PDFs calculation. The PDFs are computed using only lags in declination. The tick on left corner shows the physical scale. \textit{(Lower panel).} PDFs with different lags applied: $\delta \mathbf{r}$ = 2 (bold solid line), 5 (dotted line),  10 (thin solid line) and  15 (dashed line). The red line shows a Gaussian fit of $\delta \mathbf{r}$ = 15 PDF.}
    \label{fig:pdf_ridge}
\end{figure}

To perform the centroid analysis by taking lags along the ridge, we rotated the centroids map to align the west ridge with vertical axis and subtracted a velocity trend due to the systematic flow obtained from the CH peak velocities (see Fig. \ref{fig:diagonal_ch}) as an independent method to perform the detrending. Finally, we computed the $^{12}$CO PDF along the ridge only to maximize the range of possible lags.

Fig. \ref{fig:pdf_ridge} shows the result of this procedure. In the upper panel is highlighted the map sub-section used for the analysis: only the spectra inside the black rectangle were used in the PDF construction. For each lag there are at least 18 centroid samples, that is the width of the selected region. Outside this region the S/N is too low to perform a statistically significant analysis. In the lower panel are reported some PDFs computed for different lags.

The width of PDFs increases with lag, indicating that the velocity fluctuations decorrelate. We, consequently, estimate the correlation length by the lag where the PDF is completely relaxed.\footnote{For the observations, the integration cannot be performed between infinite velocity limits. It is customary to window  each spectrum as tightly as possible, with the line centered and removing the channels at the extremes where there is no signal.}  We calculated the dispersion for different lags (1 to 35).  Assuming a cloud distance of 93 pc, the maximum lag corresponds to about 0.7 pc. In Fig. \ref{fig:pdf_ridge} a Gaussian fit to the PDF is shown for $\delta\mathbf{r}$ = 15. As remarked before, the individual profiles in the FCRAO data cannot be used to search for intermittency because the wings are weaker than the noise level. The Gaussian fit shows that the centroids PDF is underdispersed relative to the widths of the individual profiles (interpreted as line of sight integrated PDFs, see next section) and to an uncorrelated process. The dispersion relaxes at about 0.4 pc (20 lags), which we identify as the correlation length, that is an essential signature of a turbulent flow (e.g., \cite{2000tufl.book.....P}).

\subsection{Single profiles as line of sight PDF}\label{sub:single_pdf}
An alternate approach is to treat each line profile as a sample of the turbulent velocity distribution along the line of sight, weighted by the gas emissivity. This requires high fidelity profiles but we have those in the OSO observations.  If the mean flow velocity is removed, each profile is then a PDF of the $\delta$v around the centroid. This is a robust procedure for revealing departures from Gaussian distribution and does not require detrending. 

\begin{figure}
  \centering
  \includegraphics[width=1.015\hsize]{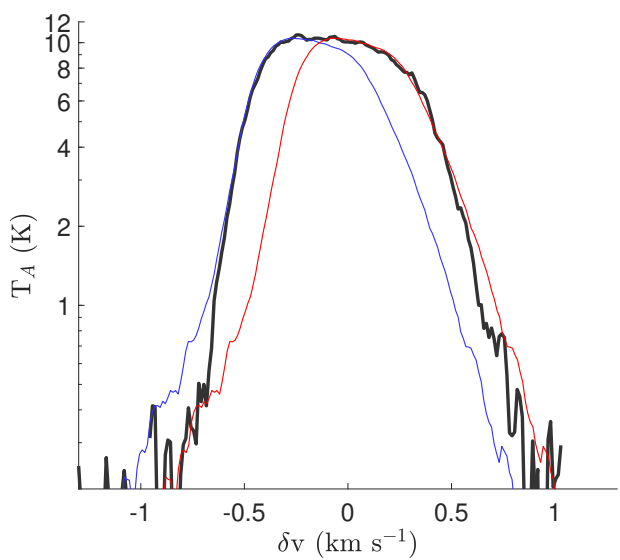}
     \caption{Comparison between \element[][12]CO composite profile in MAP1 (black thick line) and \element[][12]{CO}{} composite profile in MAP2 (blue thin line). The left wings of both profiles are almost coincident, except where $\delta$v < -0.6 km s$^{-1}$. If we apply a velocity shift of $\sim$ +0.2 km s$^{-1}$ on MAP2 composite profile (red thin line) it is possible to match also the right wings of both profiles. Again there is a small departure at high $\delta$v.}
         \label{fig:pdf_line}
\end{figure}

To show this, we compare the wings of a composite profile in MAP2 and  MAP1. We averaged all 81 profiles in each map to increase the S/N and enhance the departures from Gaussian distribution at high $\delta$v and low amplitude.  The mean MAP2 profile is a  bit narrower than that of MAP1, but the wings are nearly identical. Fig. \ref{fig:pdf_line} shows the results of our analysis: the MAP1 composite profile is shown in black thick line, where MAP2 is shown in thin blue line. As said before, the MAP2 composite profile is relatively narrower respect to MAP1 profile, so the right wings do not match. However if we shift the MAP2 profile by +0.2 km s$^{-1}$ (red thin line) we obtained a nearly perfect match with MAP1 right wing. The departures at |$\Delta$ v| >  0.6 km s$^{-1}$ in both wings (more evident on the left side) is perhaps an indication of a different intermittency scale for the two regions. This shift is independently determined but the same as the bulk CH centroid velocity difference between MAP1 and MAP2 (see Fig. \ref{fig:diagonal_ch}). 

We did not perform any functional fit to the profiles as we did not assume any velocity distribution along the line of sight. The remarkable similarity of the MAP1 and MAP2 profiles is a indication that we are mapping the same turbulent regime. Differences in far wings at low $T_A$ may indicate that the intermittency scale is changing across the filament. The CH to \element[][12]{CO}{} velocity difference between MAP1 and MAP2 is remarkably similar to the shift we have to apply to match the right wings of the \element[][12]{CO}{} profiles, as shown in Fig. \ref{fig:pdf_line}: since the right side of \element[][12]{CO}{} is tracing the outer optically thin gas (also traced  by CH) it is not surprising that the velocity shear is the same for both species.

\subsection{Structure functions}

The structure function (SF) characterizes the spatial correlation of observational proxies of flow properties, such as the integrated line intensity, velocity dispersion and skew of the profile (\cite{1985ApJ...295..466K}, \cite{1985ApJ...295..479D}). Because turbulence is a correlated process (at least at small scales, where the single turbulent filaments can be spatially resolved) SF are a useful way to represent the dynamics (see \cite{1991JFM...225....1V}, \cite{2003ApJ...583..308P} and \cite{2000tufl.book.....P}). For a map of some quantity $A$, the SF of order $p$ is
\begin{equation}
    S_p(A,\mathbf{r}) = \langle \ \left| A(\mathbf{r}) - A(\mathbf{r}+\delta \mathbf{r}) \right|^p \ \rangle
    \label{eq:sf}
\end{equation}
The angle bracket $\langle \cdot \rangle$ indicates a spatial mean over the map and $\delta \mathbf{r}$ is all possible lags inside the map. We subtracted the mean velocity (calculated as the centroid of averaged profile over the whole cloud) from each map before the SF computation.

\begin{figure}
  \centering
  \includegraphics[width=\hsize]{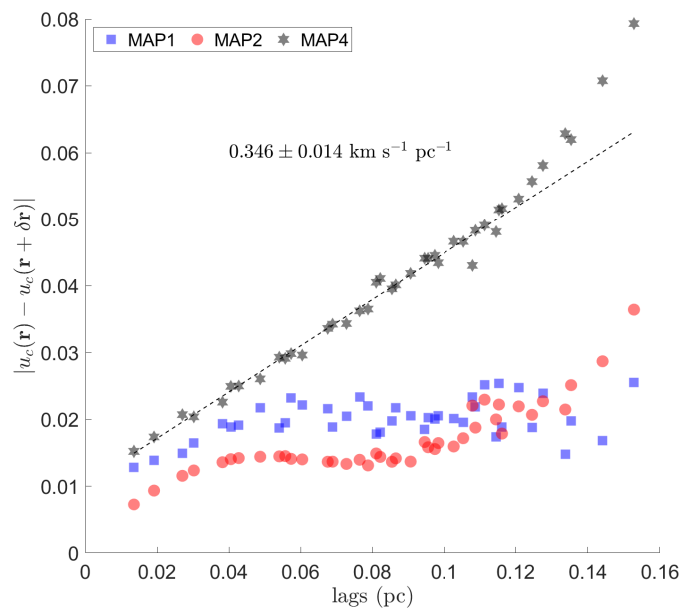}
     \caption{First structure function (SF$_1$$(u_c,\delta \mathbf{r})$) of velocity centroids in MAP1, MAP2 and MAP4 in function of the lags. MAP4 shows a linear relation over spatial scale of $\sim$ 1 pc. The dashed line is the best fit for MAP4 structure function between 0.02 and 0.12 pc.}
         \label{fig:sf_cent}
\end{figure}

\begin{figure}
  \centering
  \includegraphics[width=\hsize]{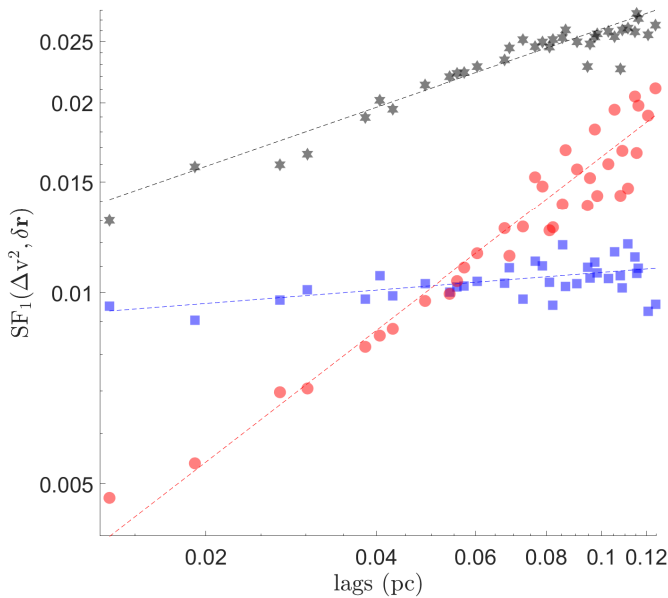}
     \caption{First structure function of the variances as a  function of the lags. Same legend of Fig. \ref{fig:sf_cent}. We removed the points between 0.13 and 0.16 pc of lags due to excessive scatter and low S/N. The slopes of the best fit (dashed lines) are for MAP1 (blue squares) $0.07 \pm 0.09$, for MAP2 (red circles) $0.69 \pm 0.05$ and $0.31 \pm 0.03$ for MAP4 (black stars).}
         \label{fig:dispersion_lags}
\end{figure}

We computed the structure functions both for the velocity centroids and for the variance. The variance\footnote{We  reserve $\sigma^2$ only for the variance of a Gaussian.} is defined as 
\begin{equation}
    \Delta \mathrm{v}^2  = \int_{line} T_A(u)\left(u-u_c \right)^2 \ du \ \Big / \int_{line} T_A(u) \ du \ \ \ .
    \label{eq:disp}
\end{equation}
\noindent
Figure \ref{fig:sf_cent} shows the first structure function of velocity centroids in MAP1, MAP2 and MAP4. In MAP1 the increments of velocity centroids is nearly independent of lag, indicating that the region is  coherent. In MAP2 there is a change between 0.06 and 0.08 pc, perhaps due to a change in gas kinematics. The structure function for MAP4, the dashed line, is linear over a range of 1 pc.

\begin{figure}
  \centering
  \includegraphics[width=\hsize]{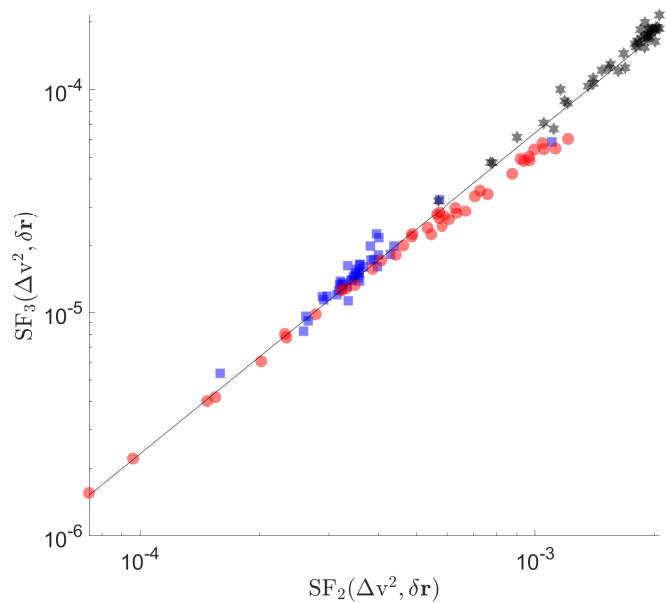}
     \caption{Relation between SF$_2(\Delta \mathrm{v}^2,\delta \mathbf{r}$) and SF$_3(\Delta \mathrm{v}^2,\delta \mathbf{r}$) in MAP1, 2 and 4 (with the same legend of Fig. \ref{fig:sf_cent}). The slope of the best fit is $1.44 \pm 0.02$.}
         \label{fig:kolmogorov}
\end{figure}

Fig. \ref{fig:dispersion_lags} shows $S_2(\delta \mathbf{r})$ for the variance ($\Delta$v$^2$) in MAP1,2 and 4, corresponding to 9 lags and using a total of 243 points. The signal in MAP3 is too weak for this procedure. In MAP1, the central velocity of each line is nearly independent of position and the variance is constant throughout the map.\footnote{ The \element[][13]CO data exhibit significant scatter, due to lower S/N in each spectrum.} The profile correlation indicates that MAP1 is a single coherent structure. In MAP2 the centroids and variances are linearly correlated with lags, but do not individually display the exponents of the SFs expected for the Kolmogorov hypothesis, see Fig. \ref{fig:kolmogorov}. This is similar to the numerical simulations by \cite{1991JFM...225....1V}.  A power law relation between structure functions of different order is, however, a general result of the similarity assumption that the energy transfer rate is independent of length scale, although  Fig. \ref{fig:kolmogorov} shows that the dissipation rate seems to be  nearly constant and scale independent along the west ridge.

For incompressible isotropic turbulence we expect Kolmogorov-like slopes, that is $2/3$ for  $p = 2$, but none of the exponents obtained from the OSO data is incompatible for this value. Several effects conspire to produce deviations in numerical simulations.  For instance, the turbulent field may be anisotropic, radiative losses may dominate the dissipation instead of  viscosity,  and the cascade have a more complex dependence on scale than the incompressible case (e.g., multifractal scaling, see \cite{1995tlan.book.....F}), or indicate the turbulent regime in MBM 40 is compressible (e.g., \cite{2013JFM...729R...1K}, \cite{2019A&A...630A..97C}).

\subsection{Kinematics and geometry of the flows}
Velocity gradients have been interpreted as accelerations resulting from large scale driving and systematic shearing.  In some cases (e.g. \cite{2002ApJ...565.1050S}) the derived gradients exceed 10 km s$^{-1}$pc$^{-1}$, suggesting strongly supersonic differential motions and possibly internal shocks. There is, however, an alternate interpretation of the P-V diagrams, such as those shown in Fig. \ref{fig:three_shear}, that the shears are produced by line of sight projections of a flow rather than accelerations.  For example, if the large scale flow has a constant streaming speed v$_s$, gradients can result from changes in the radial (projected) velocity $\mathrm{v}_{\mathrm{rad}}= \mathbf{v_s}\cdot \hat{\mathbf{s}}$, where $\hat{\mathbf s}$ is the unit vector along the line of sight. In this construction, gradients in $\mathrm{v}_{\rm rad}$ are actually the writhing and twisting of a filament. In other words, the gradients can also be produced topologically instead of dynamically. This is particularly intriguing because MBM 40 is not self-gravitating and the molecular gas is just tracer of the environmental flows.

The CH P-V diagram is consistent with a contorted flow. For instance, the multiple  components seen in some profiles could be produced by  intersections with a distorted filament along a line of sight, as sketched in Fig. \ref{fig:3d_struct} and discussed below. Thus, the line profile decomposition shown in Fig. \ref{fig:pdf_line} is explained if the turbulence producing the individual line profiles is approximately uniform along the filament. Our reconstruction assumes a constant volume emissivity, consistent with the near constancy of the dust temperature we found previously  (\cite{2022A&A...668L...9M}), and a constant streaming speed, although neither condition is strictly necessary for interpreting the profiles.

\subsection{Possible topological structure of the longitudinal shear}\label{subsec:long_shear}
In most interstellar environments it is not possible to recover the spatial structure of the object because it presents only a projected surface, and the spectroscopically derived velocity information is only linked to the third dimension for axially or spherically symmetric flows. MBM 40 shows complex line profiles, with at least two different velocity components that can be either superimposed  independent filaments or a single twisted flow that changes  orientation relative to the line of sight or both. The composite  individual profiles and the systematic shift seen in the upper part of MAP4 in \element[][12]CO point to the second  interpretation. 

\begin{figure}
  \centering
  \includegraphics[width=1.0\hsize]{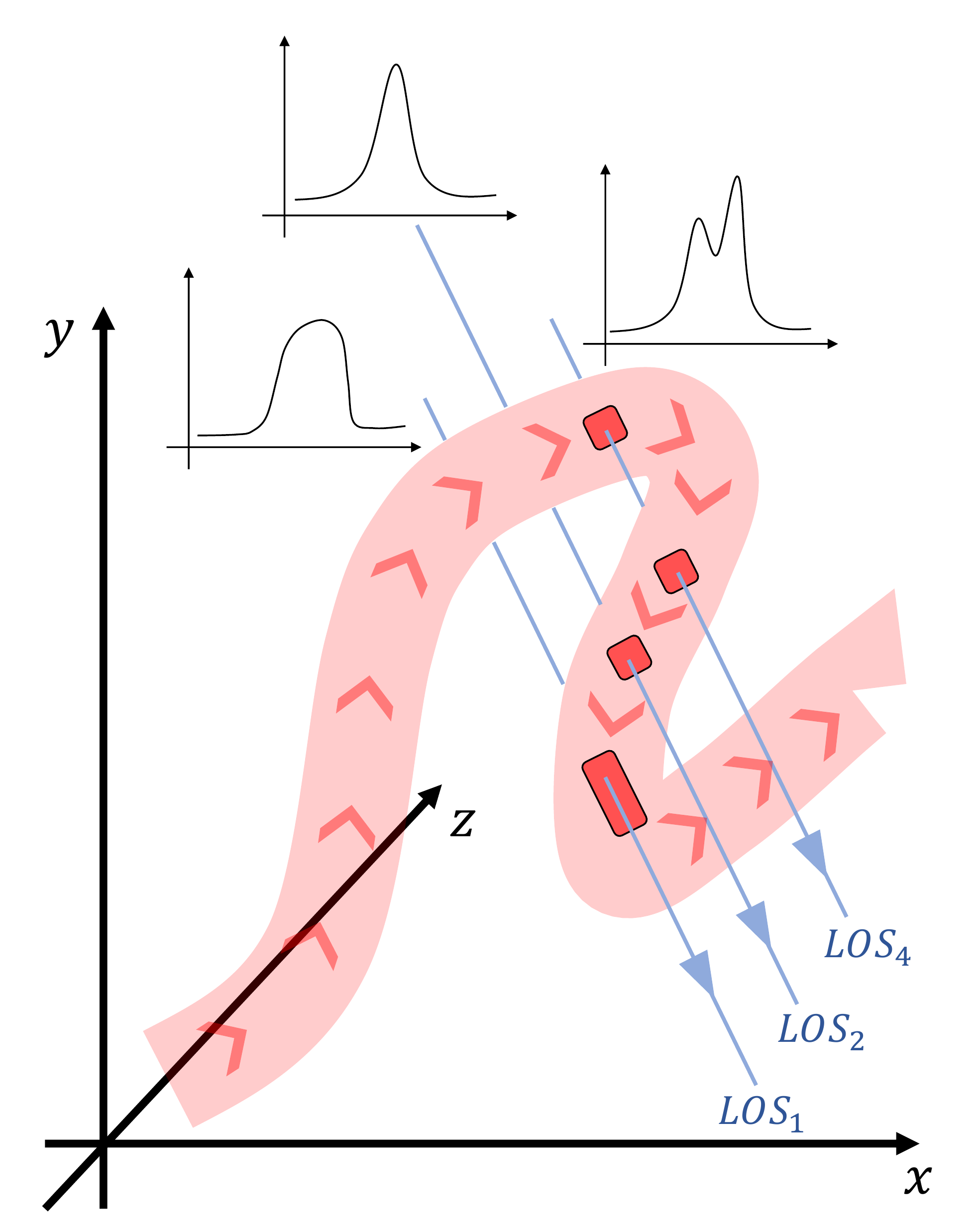}
     \caption{Possible three dimensional structure of MBM 40 with profile samples in MAP1 (line of sight 1, LOS$_1$), MAP2 (LOS$_2$), MAP4 (LOS$_4$).}
         \label{fig:3d_struct}
\end{figure}

Fig. \ref{fig:3d_struct} illustrates a possible 3D structure that consistently describes the maps. The cloud is shown as a single flow, depicted in red with flow directions indicated with red arrows. The pale blue lines are  lines of sight (LOS) shown only for MAP1, MAP2 and MAP4, and for each line of sight a sample line profile is showed. For instance, the MAP4 LOS intercepts the cloud in two different points, due to  twisting, producing the observed double-peak profile; the cloud is then bent along the $z$ direction away from us. The MAP2 LOS intercepts only a  narrow interval in radial velocity, so the profile is single and narrower. Finally, the MAP1 LOS covers a wider region of the flow due to twisting, and the line profile is wider because it spans a broader velocity range.  We therefore suggest that the line widths are also due to the combined contributions of a twisted flow and turbulent broadening, instead of two or more separated flows. If the filament thickness is comparable to the projected width of the spine, about 0.1 pc, then the inferred number density is $n \sim$ 10$^{3}$ cm$^{-3}$, compatible with the minimal density to see CO in emission, without evoking different superposed flows. 

We note that this proposed structure, assuming a constant speed along the filament, is a simplification.  There must also be a dynamical shear flow within the cloud because we see skew in single line profiles that cover a much smaller, spatially unresolved, region than the length scale over which the geometry affects the CH centroid velocities. In addition, the CH velocity centroid displacement relative to \element[][12]{CO}{} as discussed in \ref{subsec:shear} indicates the presence of a transverse shear that, however, does not compromise this picture. We cannot, however, rule out differential motions on length scales below the projected size of a beam.

\section{Conclusions}
Our aim in studying this particular diffuse high latitude cloud has been to demonstrate the complexity within even such simple interstellar structures.  Our dynamical analysis used high velocity resolution, high fidelity multi-tracer line profiles.  We propose that this cloud is embedded in a larger ($\approx$ 10 pc) scale shearing neutral hydrogen flow and is sited within a spatially and dynamically isolated extension. The molecular cloud is a three dimensional ribbon-like filament that accounts for both the large velocity gradients obtained from CO and CH tracers and the complexity of the line profiles. Treating the individual line profiles as proxies for the LOS integrated  turbulent velocity probability distributions, we find that the turbulence has a correlation scale of about 0.4 pc, roughly the size of the coherent molecular structures noted in, for instance, \cite{1999ApJ...512..761L}.  
In light of the increasing sophistication and scale of numerical simulations of the interstellar medium (\cite{1996ApJ...463..623L}, \cite{2009A&A...494..127S}, \cite{2021NatAs...5..365F}, \cite{2022MNRAS.517.5003B}), we recommend this and similar clouds as testbeds against which the models can be compared.  If such variety of processes can be found in so ordinary and comparatively tranquil environment, one can only imagine the complexity of processes when star formation drives the dynamics.

\begin{acknowledgements}
\parindent 0in
This publication makes use of data products from the {\it Wide-field Infrared Survey Explorer} ({\it WISE}), which is a joint project of the University of California, Los Angeles, and the Jet Propulsion Laboratory/California Institute of Technology, funded by the National Aeronautics and Space Administration.
\newline
This publication also make use of observations obtained with {\it Planck} (http://www.esa.int/{\it Planck}), an ESA science mission with instruments and contributions directly funded by ESA Member States, NASA, and Canada.
\newline
We acknowledge support from Onsala Space Observatory for  the  provisioning of its facilities/observational support. The Onsala Space Observatory national research infrastructure is funded through Swedish Research Council grant No 2017-00648.
\newline
The Arecibo Observatory was operated by Cornell University under a cooperative agreement with the National Science Foundation during the CH observations, and SRI International under a cooperative agreement with the National Science Foundation, and in alliance with Ana G. M\'endez-Universidad Metropolitana, and the Universities Space Research Association during the H$_2$CO observations.
\newline
The 12-meter millimeter-wave  radio telescope on Kitt Peak is operated by the Arizona Radio Observatory (ARO), Steward Observatory, University of Arizona.
\newline
We warmly thank the referee for an insightful report, Lucy Ziurys and Magnus Thomasson for their kind support of this project, Enrique Vazquez-Semadeni and John Black for enlightening discussions during our OSO run, and Ted LaRosa and Francesco Costagliola for their collaboration and friendship over the years. 
\end{acknowledgements}

\newpage

   \bibliographystyle{aa.bst} 
   \bibliography{biblio.bib} 

\begin{thebibliography}{70}
\expandafter\ifx\csname natexlab\endcsname\relax\def\natexlab#1{#1}\fi

\bibitem[{{Beattie} {et~al.}(2022){Beattie}, {Mocz}, {Federrath}, \&
  {Klessen}}]{2022MNRAS.517.5003B}
{Beattie}, J.~R., {Mocz}, P., {Federrath}, C., \& {Klessen}, R.~S. 2022,
  \mnras, 517, 5003

\bibitem[{{Belitsky} {et~al.}(2015){Belitsky}, {Lapkin}, {Fredrixon}, {Sundin},
  {Helldner}, {Pettersson}, {Ferm}, {Pantaleev}, {Billade}, {Bergman},
  {Olofsson}, {Lerner}, {Strandberg}, {Whale}, {Pavolotsky}, {Flygare},
  {Olofsson}, \& {Conway}}]{2015A&A...580A..29B}
{Belitsky}, V., {Lapkin}, I., {Fredrixon}, M., {et~al.} 2015, \aap, 580, A29

\bibitem[{{Brunt} \& {Heyer}(2013)}]{2013MNRAS.433..117B}
{Brunt}, C.~M. \& {Heyer}, M.~H. 2013, \mnras, 433, 117

\bibitem[{{Chastain} {et~al.}(2010){Chastain}, {Cotten}, \&
  {Magnani}}]{2010AJ....139..267C}
{Chastain}, R.~J., {Cotten}, D., \& {Magnani}, L. 2010, \aj, 139, 267

\bibitem[{{Chira} {et~al.}(2019){Chira}, {Ib{\'a}{\~n}ez-Mej{\'\i}a}, {Mac
  Low}, \& {Henning}}]{2019A&A...630A..97C}
{Chira}, R.~A., {Ib{\'a}{\~n}ez-Mej{\'\i}a}, J.~C., {Mac Low}, M.~M., \&
  {Henning}, T. 2019, \aap, 630, A97

\bibitem[{{Cotten} \& {Magnani}(2013)}]{2013MNRAS.436.1152C}
{Cotten}, D.~L. \& {Magnani}, L. 2013, \mnras, 436, 1152

\bibitem[{{Cotten} {et~al.}(2012){Cotten}, {Magnani}, {Wennerstrom}, {Douglas},
  \& {Onello}}]{2012AJ....144..163C}
{Cotten}, D.~L., {Magnani}, L., {Wennerstrom}, E.~A., {Douglas}, K.~A., \&
  {Onello}, J.~S. 2012, \aj, 144, 163

\bibitem[{{Dame} \& {Thaddeus}(2022)}]{2022ApJS..262....5D}
{Dame}, T.~M. \& {Thaddeus}, P. 2022, \apjs, 262, 5

\bibitem[{{Dickman} \& {Kleiner}(1985)}]{1985ApJ...295..479D}
{Dickman}, R.~L. \& {Kleiner}, S.~C. 1985, \apj, 295, 479

\bibitem[{{Draine}(2011)}]{2011piim.book.....D}
{Draine}, B.~T. 2011, {Physics of the Interstellar and Intergalactic Medium}

\bibitem[{{Elmegreen} \& {Scalo}(2004)}]{2004ARA&A..42..211E}
{Elmegreen}, B.~G. \& {Scalo}, J. 2004, \araa, 42, 211

\bibitem[{{Esquivel} \& {Lazarian}(2005)}]{2005ApJ...631..320E}
{Esquivel}, A. \& {Lazarian}, A. 2005, \apj, 631, 320

\bibitem[{{Falceta-Gon{\c{c}}alves} {et~al.}(2014){Falceta-Gon{\c{c}}alves},
  {Kowal}, {Falgarone}, \& {Chian}}]{2014NPGeo..21..587F}
{Falceta-Gon{\c{c}}alves}, D., {Kowal}, G., {Falgarone}, E., \& {Chian},
  A.~C.~L. 2014, Nonlinear Processes in Geophysics, 21, 587

\bibitem[{{Falgarone} {et~al.}(1998){Falgarone}, {Panis}, {Heithausen},
  {Perault}, {Stutzki}, {Puget}, \& {Bensch}}]{1998A&A...331..669F}
{Falgarone}, E., {Panis}, J.~F., {Heithausen}, A., {et~al.} 1998, \aap, 331,
  669

\bibitem[{{Falgarone} {et~al.}(2009){Falgarone}, {Pety}, \&
  {Hily-Blant}}]{2009A&A...507..355F}
{Falgarone}, E., {Pety}, J., \& {Hily-Blant}, P. 2009, \aap, 507, 355

\bibitem[{{Falgarone} \& {Phillips}(1990)}]{1990ApJ...359..344F}
{Falgarone}, E. \& {Phillips}, T.~G. 1990, \apj, 359, 344

\bibitem[{{Falgarone} {et~al.}(1991){Falgarone}, {Phillips}, \&
  {Walker}}]{1991ApJ...378..186F}
{Falgarone}, E., {Phillips}, T.~G., \& {Walker}, C.~K. 1991, \apj, 378, 186

\bibitem[{{Falgarone} \& {Puget}(1995)}]{1995A&A...293..840F}
{Falgarone}, E. \& {Puget}, J.~L. 1995, \aap, 293, 840

\bibitem[{{Federrath} {et~al.}(2021){Federrath}, {Klessen}, {Iapichino}, \&
  {Beattie}}]{2021NatAs...5..365F}
{Federrath}, C., {Klessen}, R.~S., {Iapichino}, L., \& {Beattie}, J.~R. 2021,
  Nature Astronomy, 5, 365

\bibitem[{{Frisch}(1995)}]{1995tlan.book.....F}
{Frisch}, U. 1995, {Turbulence. The legacy of A.N. Kolmogorov}

\bibitem[{{Hearty} {et~al.}(1999){Hearty}, {Magnani}, {Caillault},
  {Neuh{\"a}user}, {Schmitt}, \& {Stauffer}}]{1999A&A...341..163H}
{Hearty}, T., {Magnani}, L., {Caillault}, J.~P., {et~al.} 1999, \aap, 341, 163

\bibitem[{{Hennebelle} \& {Falgarone}(2012)}]{2012A&ARv..20...55H}
{Hennebelle}, P. \& {Falgarone}, E. 2012, \aapr, 20, 55

\bibitem[{{Heyer} \& {Dame}(2015)}]{2015ARA&A..53..583H}
{Heyer}, M. \& {Dame}, T.~M. 2015, \araa, 53, 583

\bibitem[{{Hily-Blant} {et~al.}(2008){Hily-Blant}, {Falgarone}, \&
  {Pety}}]{2008A&A...481..367H}
{Hily-Blant}, P., {Falgarone}, E., \& {Pety}, J. 2008, \aap, 481, 367

\bibitem[{{Imara} \& {Blitz}(2011)}]{2011ApJ...732...78I}
{Imara}, N. \& {Blitz}, L. 2011, \apj, 732, 78

\bibitem[{{Kleiner} \& {Dickman}(1984)}]{1984ApJ...286..255K}
{Kleiner}, S.~C. \& {Dickman}, R.~L. 1984, \apj, 286, 255

\bibitem[{{Kleiner} \& {Dickman}(1985)}]{1985ApJ...295..466K}
{Kleiner}, S.~C. \& {Dickman}, R.~L. 1985, \apj, 295, 466

\bibitem[{{Koda} {et~al.}(2006){Koda}, {Sawada}, {Hasegawa}, \&
  {Scoville}}]{2006ApJ...638..191K}
{Koda}, J., {Sawada}, T., {Hasegawa}, T., \& {Scoville}, N.~Z. 2006, \apj, 638,
  191

\bibitem[{{Kritsuk} {et~al.}(2013){Kritsuk}, {Wagner}, \&
  {Norman}}]{2013JFM...729R...1K}
{Kritsuk}, A.~G., {Wagner}, R., \& {Norman}, M.~L. 2013, Journal of Fluid
  Mechanics, 729, R1

\bibitem[{{Kutner} \& {Ulich}(1981)}]{1981ApJ...250..341K}
{Kutner}, M.~L. \& {Ulich}, B.~L. 1981, \apj, 250, 341

\bibitem[{{LaRosa} {et~al.}(1999){LaRosa}, {Shore}, \&
  {Magnani}}]{1999ApJ...512..761L}
{LaRosa}, T.~N., {Shore}, S.~N., \& {Magnani}, L. 1999, \apj, 512, 761

\bibitem[{{Lazarian} \& {Pogosyan}(2000)}]{2000ApJ...537..720L}
{Lazarian}, A. \& {Pogosyan}, D. 2000, \apj, 537, 720

\bibitem[{{Lee} {et~al.}(2002){Lee}, {Chung}, \& {Kim}}]{2002JKAS...35...97L}
{Lee}, Y., {Chung}, H.~S., \& {Kim}, H. 2002, Journal of Korean Astronomical
  Society, 35, 97

\bibitem[{{Lesieur}(2008)}]{2008tufl.book.....L}
{Lesieur}, M. 2008, {Turbulence in Fluids}

\bibitem[{{Lis} {et~al.}(1996){Lis}, {Pety}, {Phillips}, \&
  {Falgarone}}]{1996ApJ...463..623L}
{Lis}, D.~C., {Pety}, J., {Phillips}, T.~G., \& {Falgarone}, E. 1996, \apj,
  463, 623

\bibitem[{{Liszt} \& {Lucas}(2002)}]{2002A&A...391..693L}
{Liszt}, H. \& {Lucas}, R. 2002, \aap, 391, 693

\bibitem[{{Liszt} {et~al.}(2005){Liszt}, {Lucas}, \&
  {Pety}}]{2005IAUS..231..187L}
{Liszt}, H., {Lucas}, R., \& {Pety}, J. 2005, in Astrochemistry: Recent
  Successes and Current Challenges, ed. D.~C. {Lis}, G.~A. {Blake}, \&
  E.~{Herbst}, Vol. 231, 187--196

\bibitem[{{Mac Low} {et~al.}(1998){Mac Low}, {Klessen}, {Burkert}, \&
  {Smith}}]{1998PhRvL..80.2754M}
{Mac Low}, M.-M., {Klessen}, R.~S., {Burkert}, A., \& {Smith}, M.~D. 1998,
  \prl, 80, 2754

\bibitem[{{Magnani} {et~al.}(1985){Magnani}, {Blitz}, \&
  {Mundy}}]{1985ApJ...295..402M}
{Magnani}, L., {Blitz}, L., \& {Mundy}, L. 1985, \apj, 295, 402

\bibitem[{{Magnani} {et~al.}(1996){Magnani}, {Caillault}, {Hearty}, {Stauffer},
  {Schmitt}, {Neuhaeuser}, {Verter}, \& {Dwek}}]{1996ApJ...465..825M}
{Magnani}, L., {Caillault}, J.-P., {Hearty}, T., {et~al.} 1996, \apj, 465, 825

\bibitem[{{Magnani} {et~al.}(1993){Magnani}, {Larosa}, \&
  {Shore}}]{1993ApJ...402..226M}
{Magnani}, L., {Larosa}, T.~N., \& {Shore}, S.~N. 1993, \apj, 402, 226

\bibitem[{{Magnani} {et~al.}(1998){Magnani}, {Onello}, {Adams}, {Hartmann}, \&
  {Thaddeus}}]{1998ApJ...504..290M}
{Magnani}, L., {Onello}, J.~S., {Adams}, N.~G., {Hartmann}, D., \& {Thaddeus},
  P. 1998, \apj, 504, 290

\bibitem[{{Magnani} \& {Shore}(2017)}]{2017ASSL..442.....M}
{Magnani}, L. \& {Shore}, S.~N. 2017, {A Dirty Window}, Vol. 442

\bibitem[{{Mattila}(1986)}]{1986A&A...160..157M}
{Mattila}, K. 1986, \aap, 160, 157

\bibitem[{{McComb}(1990)}]{1990cp...book.....M}
{McComb}, W.~D. 1990, {The physics of fluid turbulence}

\bibitem[{{Miesch} \& {Bally}(1994)}]{1994ApJ...429..645M}
{Miesch}, M.~S. \& {Bally}, J. 1994, \apj, 429, 645

\bibitem[{{Miesch} {et~al.}(1999){Miesch}, {Scalo}, \&
  {Bally}}]{1999ApJ...524..895M}
{Miesch}, M.~S., {Scalo}, J., \& {Bally}, J. 1999, \apj, 524, 895

\bibitem[{{Monaci} {et~al.}(2022){Monaci}, {Magnani}, \&
  {Shore}}]{2022A&A...668L...9M}
{Monaci}, M., {Magnani}, L., \& {Shore}, S.~N. 2022, \aap, 668, L9

\bibitem[{{Ossenkopf} \& {Mac Low}(2002)}]{2002A&A...390..307O}
{Ossenkopf}, V. \& {Mac Low}, M.~M. 2002, \aap, 390, 307

\bibitem[{{Padoan} {et~al.}(2003){Padoan}, {Boldyrev}, {Langer}, \&
  {Nordlund}}]{2003ApJ...583..308P}
{Padoan}, P., {Boldyrev}, S., {Langer}, W., \& {Nordlund}, {\r{A}}. 2003, \apj,
  583, 308

\bibitem[{{Peek} {et~al.}(2018){Peek}, {Babler}, {Zheng}, {Clark}, {Douglas},
  {Korpela}, {Putman}, {Stanimirovi{\'c}}, {Gibson}, \&
  {Heiles}}]{2018ApJS..234....2P}
{Peek}, J.~E.~G., {Babler}, B.~L., {Zheng}, Y., {et~al.} 2018, \apjs, 234, 2

\bibitem[{{Peek} {et~al.}(2011){Peek}, {Heiles}, {Douglas}, {Lee}, {Grcevich},
  {Stanimirovi{\'c}}, {Putman}, {Korpela}, {Gibson}, {Begum}, {Saul},
  {Robishaw}, \& {Kr{\v{c}}o}}]{2011ApJS..194...20P}
{Peek}, J.~E.~G., {Heiles}, C., {Douglas}, K.~A., {et~al.} 2011, \apjs, 194, 20

\bibitem[{{Pety} \& {Falgarone}(2003)}]{2003A&A...412..417P}
{Pety}, J. \& {Falgarone}, E. 2003, \aap, 412, 417

\bibitem[{{Planck Collaboration III} {et~al.}(2020){Planck Collaboration III},
  {Aghanim}, {Akrami}, {Ashdown}, {Aumont}, {Baccigalupi}, {Ballardini},
  {Banday}, {Barreiro}, {Bartolo}, {Basak}, {Benabed}, {Bernard}, {Bersanelli},
  {Bielewicz}, {Bond}, {Borrill}, {Bouchet}, {Boulanger}, {Bucher}, {Burigana},
  {Calabrese}, {Cardoso}, {Carron}, {Challinor}, {Chiang}, {Colombo}, {Combet},
  {Couchot}, {Crill}, {Cuttaia}, {de Bernardis}, {de Rosa}, {de Zotti},
  {Delabrouille}, {Delouis}, {Di Valentino}, {Diego}, {Dor{\'e}}, {Douspis},
  {Ducout}, {Dupac}, {Efstathiou}, {Elsner}, {En{\ss}lin}, {Eriksen},
  {Falgarone}, {Fantaye}, {Finelli}, {Frailis}, {Fraisse}, {Franceschi},
  {Frolov}, {Galeotta}, {Galli}, {Ganga}, {G{\'e}nova-Santos}, {Gerbino},
  {Ghosh}, {Gonz{\'a}lez-Nuevo}, {G{\'o}rski}, {Gratton}, {Gruppuso},
  {Gudmundsson}, {Handley}, {Hansen}, {Henrot-Versill{\'e}}, {Herranz},
  {Hivon}, {Huang}, {Jaffe}, {Jones}, {Karakci}, {Keih{\"a}nen}, {Keskitalo},
  {Kiiveri}, {Kim}, {Kisner}, {Krachmalnicoff}, {Kunz}, {Kurki-Suonio},
  {Lagache}, {Lamarre}, {Lasenby}, {Lattanzi}, {Lawrence}, {Levrier},
  {Liguori}, {Lilje}, {Lindholm}, {L{\'o}pez-Caniego}, {Ma},
  {Mac{\'\i}as-P{\'e}rez}, {Maggio}, {Maino}, {Mandolesi}, {Mangilli},
  {Martin}, {Mart{\'\i}nez-Gonz{\'a}lez}, {Matarrese}, {Mauri}, {McEwen},
  {Melchiorri}, {Mennella}, {Migliaccio}, {Miville-Desch{\^e}nes}, {Molinari},
  {Moneti}, {Montier}, {Morgante}, {Moss}, {Mottet}, {Natoli}, {Pagano},
  {Paoletti}, {Partridge}, {Patanchon}, {Patrizii}, {Perdereau}, {Perrotta},
  {Pettorino}, {Piacentini}, {Puget}, {Rachen}, {Reinecke}, {Remazeilles},
  {Renzi}, {Rocha}, {Roudier}, {Salvati}, {Sandri}, {Savelainen}, {Scott},
  {Sirignano}, {Sirri}, {Spencer}, {Sunyaev}, {Suur-Uski}, {Tauber},
  {Tavagnacco}, {Tenti}, {Toffolatti}, {Tomasi}, {Tristram}, {Trombetti},
  {Valiviita}, {Vansyngel}, {Van Tent}, {Vibert}, {Vielva}, {Villa},
  {Vittorio}, {Wandelt}, {Wehus}, \& {Zonca}}]{2020A&A...641A...3P}
{Planck Collaboration III}, {Aghanim}, N., {Akrami}, Y., {et~al.} 2020, \aap,
  641, A3

\bibitem[{{Planck Collaboration IV} {et~al.}(2020){Planck Collaboration IV},
  {Akrami}, {Ashdown}, {Aumont}, {Baccigalupi}, {Ballardini}, {Banday},
  {Barreiro}, {Bartolo}, {Basak}, {Benabed}, {Bersanelli}, {Bielewicz}, {Bond},
  {Borrill}, {Bouchet}, {Boulanger}, {Bucher}, {Burigana}, {Calabrese},
  {Cardoso}, {Carron}, {Casaponsa}, {Challinor}, {Colombo}, {Combet}, {Crill},
  {Cuttaia}, {de Bernardis}, {de Rosa}, {de Zotti}, {Delabrouille}, {Delouis},
  {Di Valentino}, {Dickinson}, {Diego}, {Donzelli}, {Dor{\'e}}, {Ducout},
  {Dupac}, {Efstathiou}, {Elsner}, {En{\ss}lin}, {Eriksen}, {Falgarone},
  {Fernandez-Cobos}, {Finelli}, {Forastieri}, {Frailis}, {Fraisse},
  {Franceschi}, {Frolov}, {Galeotta}, {Galli}, {Ganga}, {G{\'e}nova-Santos},
  {Gerbino}, {Ghosh}, {Gonz{\'a}lez-Nuevo}, {G{\'o}rski}, {Gratton},
  {Gruppuso}, {Gudmundsson}, {Handley}, {Hansen}, {Helou}, {Herranz},
  {Hildebrandt}, {Huang}, {Jaffe}, {Karakci}, {Keih{\"a}nen}, {Keskitalo},
  {Kiiveri}, {Kim}, {Kisner}, {Krachmalnicoff}, {Kunz}, {Kurki-Suonio},
  {Lagache}, {Lamarre}, {Lasenby}, {Lattanzi}, {Lawrence}, {Le Jeune},
  {Levrier}, {Liguori}, {Lilje}, {Lindholm}, {L{\'o}pez-Caniego}, {Lubin},
  {Ma}, {Mac{\'\i}as-P{\'e}rez}, {Maggio}, {Maino}, {Mandolesi}, {Mangilli},
  {Marcos-Caballero}, {Maris}, {Martin}, {Mart{\'\i}nez-Gonz{\'a}lez},
  {Matarrese}, {Mauri}, {McEwen}, {Meinhold}, {Melchiorri}, {Mennella},
  {Migliaccio}, {Miville-Desch{\^e}nes}, {Molinari}, {Moneti}, {Montier},
  {Morgante}, {Natoli}, {Oppizzi}, {Pagano}, {Paoletti}, {Partridge}, {Peel},
  {Pettorino}, {Piacentini}, {Polenta}, {Puget}, {Rachen}, {Reinecke},
  {Remazeilles}, {Renzi}, {Rocha}, {Roudier}, {Rubi{\~n}o-Mart{\'\i}n},
  {Ruiz-Granados}, {Salvati}, {Sandri}, {Savelainen}, {Scott}, {Seljebotn},
  {Sirignano}, {Spencer}, {Suur-Uski}, {Tauber}, {Tavagnacco}, {Tenti},
  {Thommesen}, {Toffolatti}, {Tomasi}, {Trombetti}, {Valiviita}, {Van Tent},
  {Vielva}, {Villa}, {Vittorio}, {Wandelt}, {Wehus}, {Zacchei}, \&
  {Zonca}}]{2020A&A...641A...4P}
{Planck Collaboration IV}, {Akrami}, Y., {Ashdown}, M., {et~al.} 2020, \aap,
  641, A4

\bibitem[{{Pope}(2000)}]{2000tufl.book.....P}
{Pope}, S.~B. 2000, {Turbulent Flows}

\bibitem[{{Reach} {et~al.}(2015){Reach}, {Heiles}, \&
  {Bernard}}]{2015ApJ...811..118R}
{Reach}, W.~T., {Heiles}, C., \& {Bernard}, J.-P. 2015, \apj, 811, 118

\bibitem[{{Rydbeck} {et~al.}(1976){Rydbeck}, {Kollberg}, {Hjalmarson}, {Sume},
  {Ellder}, \& {Irvine}}]{1976ApJS...31..333R}
{Rydbeck}, O.~E.~H., {Kollberg}, E., {Hjalmarson}, A., {et~al.} 1976, \apjs,
  31, 333

\bibitem[{{Sakamoto}(2002)}]{2002ApJ...565.1050S}
{Sakamoto}, S. 2002, \apj, 565, 1050

\bibitem[{{Schlegel} {et~al.}(1998){Schlegel}, {Finkbeiner}, \&
  {Davis}}]{1998ApJ...500..525S}
{Schlegel}, D.~J., {Finkbeiner}, D.~P., \& {Davis}, M. 1998, \apj, 500, 525

\bibitem[{{Schmidt} {et~al.}(2009){Schmidt}, {Federrath}, {Hupp}, {Kern}, \&
  {Niemeyer}}]{2009A&A...494..127S}
{Schmidt}, W., {Federrath}, C., {Hupp}, M., {Kern}, S., \& {Niemeyer}, J.~C.
  2009, \aap, 494, 127

\bibitem[{{Shore} {et~al.}(2006){Shore}, {Larosa}, {Chastain}, \&
  {Magnani}}]{2006A&A...457..197S}
{Shore}, S.~N., {Larosa}, T.~N., {Chastain}, R.~J., \& {Magnani}, L. 2006,
  \aap, 457, 197

\bibitem[{{Shore} {et~al.}(2003){Shore}, {Magnani}, {LaRosa}, \&
  {McCarthy}}]{2003ApJ...593..413S}
{Shore}, S.~N., {Magnani}, L., {LaRosa}, T.~N., \& {McCarthy}, M.~N. 2003,
  \apj, 593, 413

\bibitem[{Tennekes {et~al.}(1972)Tennekes, Lumley, Lumley,
  {et~al.}}]{tennekes1972first}
Tennekes, H., Lumley, J.~L., Lumley, J.~L., {et~al.} 1972, A first course in
  turbulence (MIT press)

\bibitem[{{Tucker} {et~al.}(1970){Tucker}, {Tomasevich}, \&
  {Thaddeus}}]{1970ApJ...161L.153T}
{Tucker}, K.~D., {Tomasevich}, G.~R., \& {Thaddeus}, P. 1970, \apjl, 161, L153

\bibitem[{{van der Tak} {et~al.}(2007){van der Tak}, {Black}, {Sch{\"o}ier},
  {Jansen}, \& {van Dishoeck}}]{2007A&A...468..627V}
{van der Tak}, F.~F.~S., {Black}, J.~H., {Sch{\"o}ier}, F.~L., {Jansen}, D.~J.,
  \& {van Dishoeck}, E.~F. 2007, \aap, 468, 627

\bibitem[{{van Dishoeck} \& {Black}(1988)}]{1988ApJ...334..771V}
{van Dishoeck}, E.~F. \& {Black}, J.~H. 1988, \apj, 334, 771

\bibitem[{{Vincent} \& {Meneguzzi}(1991)}]{1991JFM...225....1V}
{Vincent}, A. \& {Meneguzzi}, M. 1991, Journal of Fluid Mechanics, 225, 1

\bibitem[{{Wright} {et~al.}(2010){Wright}, {Eisenhardt}, {Mainzer}, {Ressler},
  {Cutri}, {Jarrett}, {Kirkpatrick}, {Padgett}, {McMillan}, {Skrutskie},
  {Stanford}, {Cohen}, {Walker}, {Mather}, {Leisawitz}, {Gautier}, {McLean},
  {Benford}, {Lonsdale}, {Blain}, {Mendez}, {Irace}, {Duval}, {Liu}, {Royer},
  {Heinrichsen}, {Howard}, {Shannon}, {Kendall}, {Walsh}, {Larsen}, {Cardon},
  {Schick}, {Schwalm}, {Abid}, {Fabinsky}, {Naes}, \&
  {Tsai}}]{2010AJ....140.1868W}
{Wright}, E.~L., {Eisenhardt}, P. R.~M., {Mainzer}, A.~K., {et~al.} 2010, \aj,
  140, 1868

\bibitem[{Zucker {et~al.}(2019)Zucker, Speagle, Schlafly, Green, Finkbeiner,
  Goodman, \& Alves}]{zucker2019}
Zucker, C., Speagle, J.~S., Schlafly, E.~F., {et~al.} 2019, ApJ, 879, 125

\end{thebibliography}

\begin{appendix}
\section{Summary of observations}\label{app:summary}
\begin{figure*}
  \centering
       \includegraphics[scale=0.37,bb=0 10 1500 750,clip]{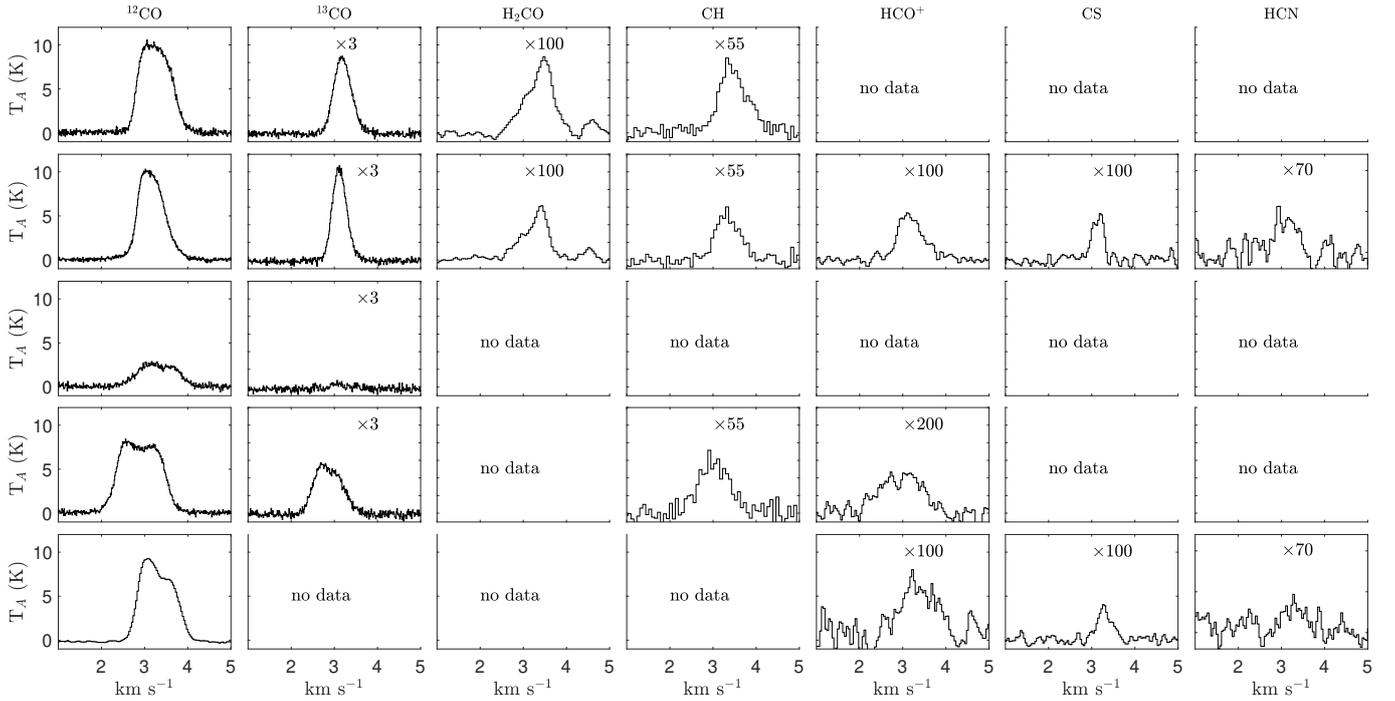}
      \caption{Sample of all observations (from top: MAP1, 2, 3, 4 and Z). Where more than one position were observed, the average throughout the map is shown. H$_2$CO is seen in absorption, so we inverted the lines (T$_A$ $\longrightarrow$ $-$T$_A$). Using the ARO telescope, we made also a deep integration in the central point of MAP2 in C$_3$H$_2$ ($2_{1,2}-1_{0,1}$, 85.339 GHz), without any detection.}
         \label{fig:summary}
\end{figure*}

In Tab. \ref{tab:summary} is shown the observations summary, with centroids, dispersions and peak temperature of each transition observed in MBM40. In Fig. \ref{fig:summary} is reported a collage of all molecular observations in various MAPs of MBM 40. Some molecule lines were multiplied by an integer factor to ease the comparison with stronger transitions.

\begin{table}
\caption{Summary of observations}             
\label{tab:summary}      
\centering                          
\begin{tabular}{c c c c c}        
\hline\hline                 
map & molecule & $u_c$ (km s$^{-1}$) & $\Delta \mathrm{v}$ (km s$^{-1}$) & T$_{max}$ (K) \\
\hline                        
MAP1 & \element[][12]CO & 3.26 & 0.33 & 9.92 \\
     & \element[][13]CO & 3.19 & 0.19 & 2.87 \\
     & H$_2$CO          & 3.43 & 0.30 & 0.08 \\
     & CH               & 3.49 & 0.32 & 0.13 \\
\hline
MAP2 & \element[][12]CO & 3.18 & 0.29 & 10.11 \\
     & \element[][13]CO & 3.10 & 0.16 & 3.31 \\
     & H$_2$CO          & 3.33 & 0.29 & 0.06 \\
     & CH               & 3.36 & 0.27 & 0.09 \\
     & HCO$^+$          & 3.21 & 0.28 & 0.05 \\
     & CS               & 3.14 & 0.13 & 0.05 \\
     & HCN              & 3.09 & 0.23 & 0.06 \\
\hline
MAP3 & \element[][12]CO & 3.26 & 0.44 & 2.69 \\
     & \element[][13]CO & -    & -    & -    \\
\hline
MAP4 & \element[][12]CO & 2.90 & 0.40 & 8.18 \\
     & \element[][13]CO & 2.88 & 0.33 & 1.88 \\
     & CH               & 3.05 & 0.35 & 0.11 \\
     & HCO$^+$          & 2.94 & 0.45 & 0.02 \\
\hline
MAPZ & \element[][12]CO & 3.29 & 0.23 & 6.18 \\
     & HCO$^+$          & 3.31 & 0.36 & 0.05 \\
     & CS               & 3.27 & 0.17 & 0.03 \\
     & HCN              & -    & -    & -    \\
\hline\hline                                   
\end{tabular}
\end{table}

\section{\element[][12]{CO} and \element[][13]{CO} OSO maps}
\begin{figure*}
   \centering
            \includegraphics[scale=0.52,bb=40 10 1100 650,clip]{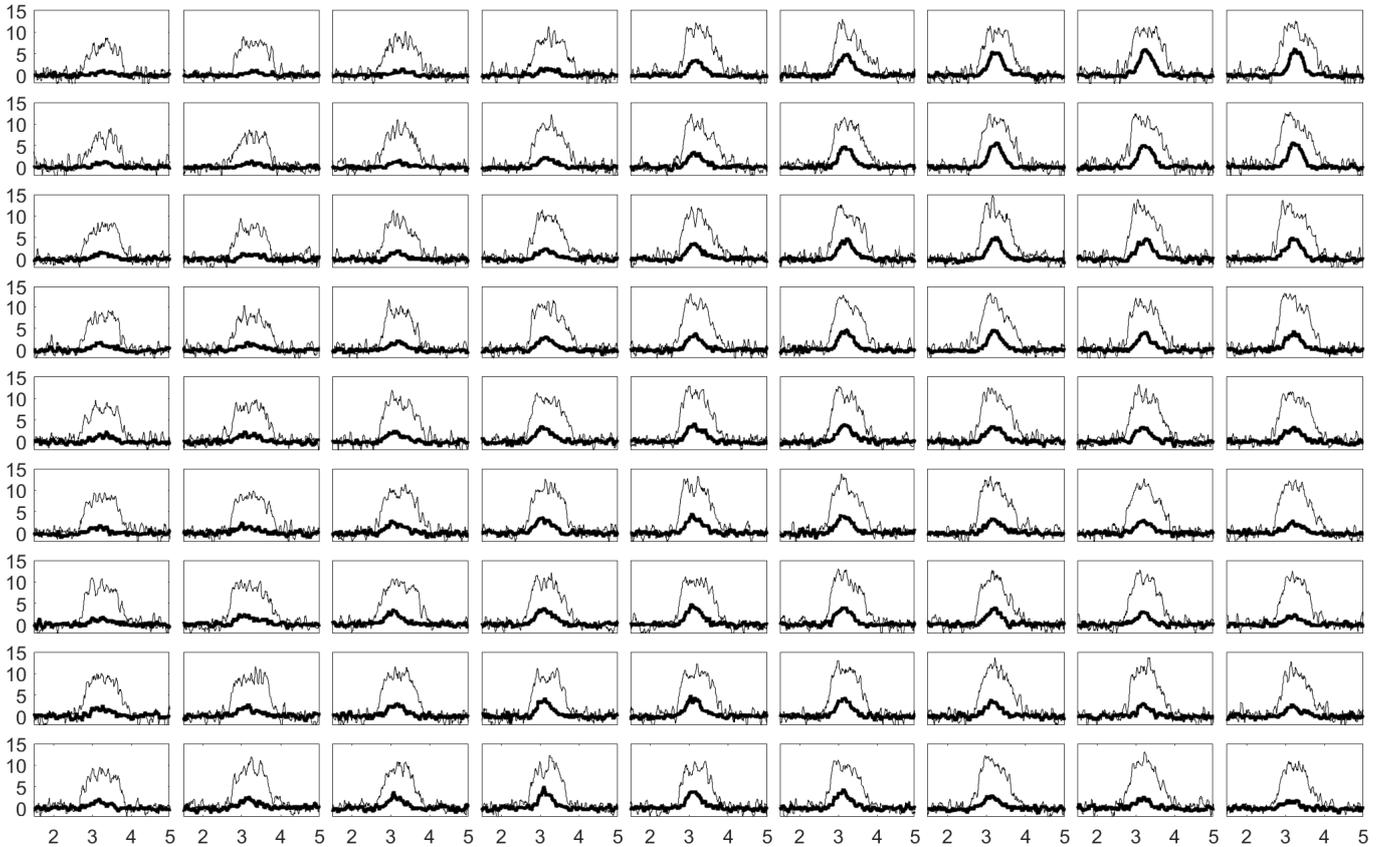}
      \caption{\element[][12]CO (thin line) and \element[][13]CO (thick line) MAP1 taken with Onsala telescope. Antenna temperature T$_A$ in K and v$_{LSR}$ in km s$^{-1}$.}
         \label{fig:map1oso}
   \end{figure*}

\begin{figure*}
    \centering
            \includegraphics[scale=0.52,bb=40 10 1100 630,clip]{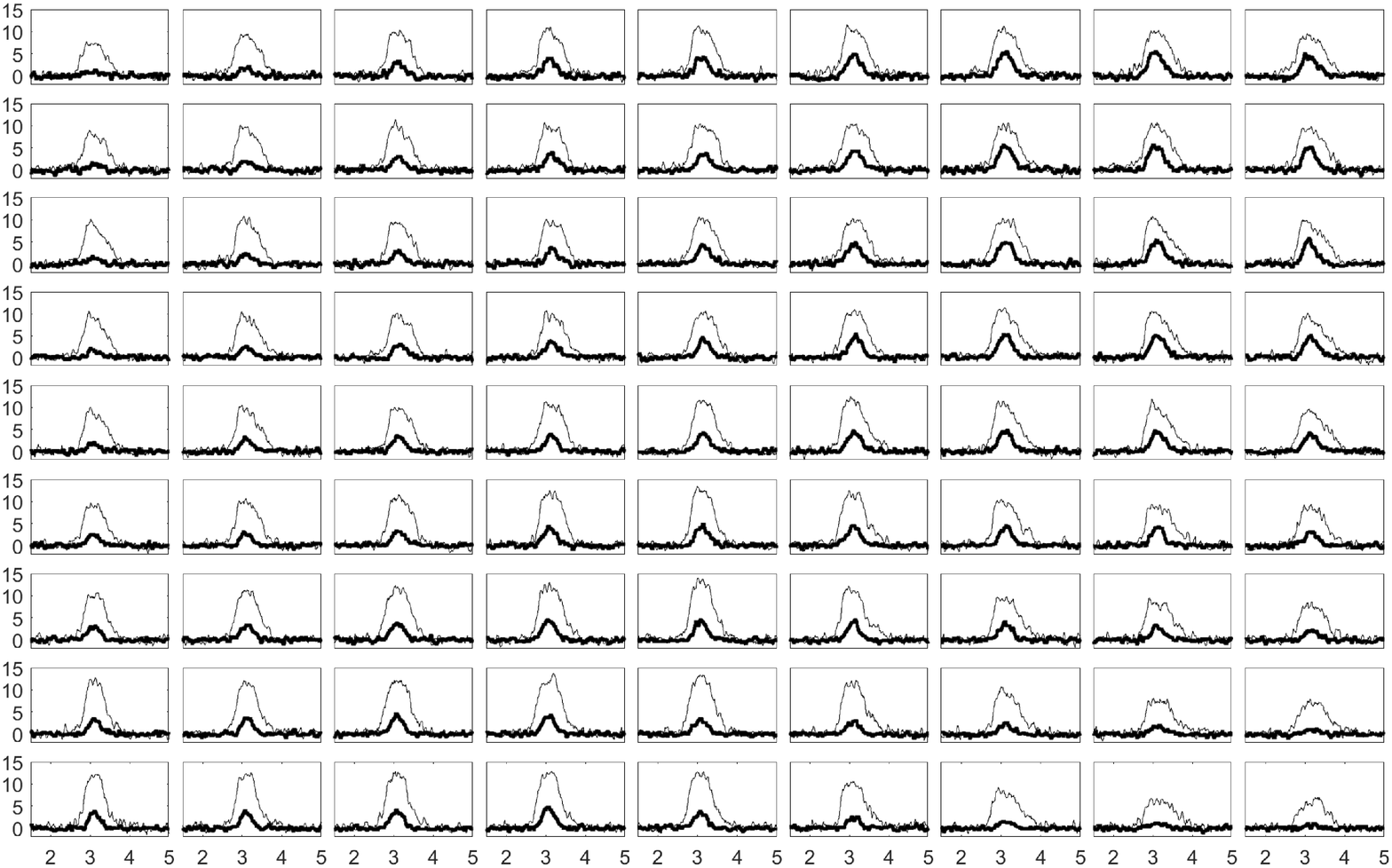}
      \caption{Same of \ref{fig:map1oso} but in MAP2.}
         \label{fig:map2oso}
   \end{figure*}

\begin{figure*}
    \centering
            \includegraphics[scale=0.52,bb=40 10 1100 630,clip]{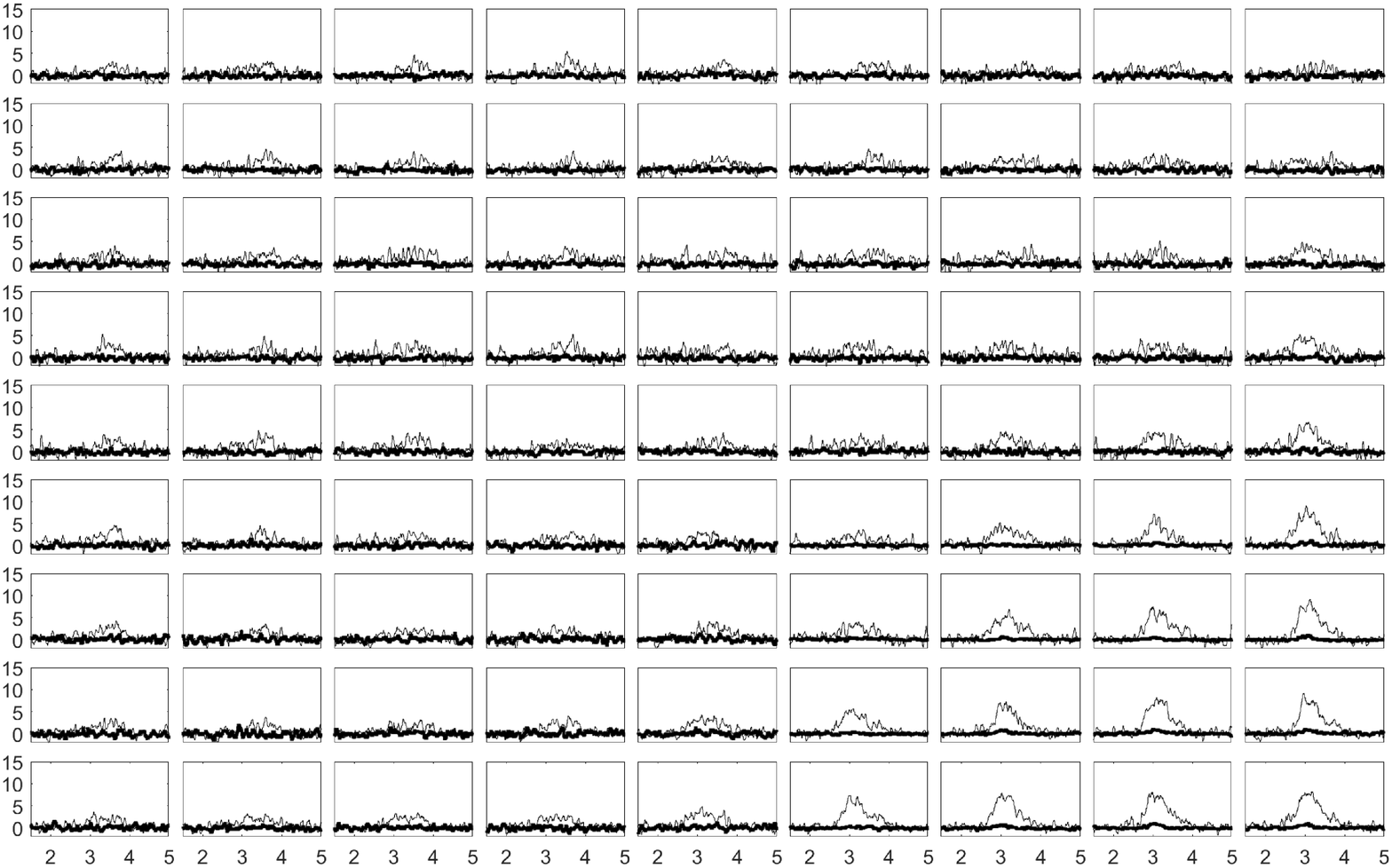}
      \caption{Same of \ref{fig:map1oso} but in MAP3.}
         \label{fig:map3oso}
   \end{figure*}

\begin{figure*}
    \centering
            \includegraphics[scale=0.52,bb=40 10 1100 650,clip]{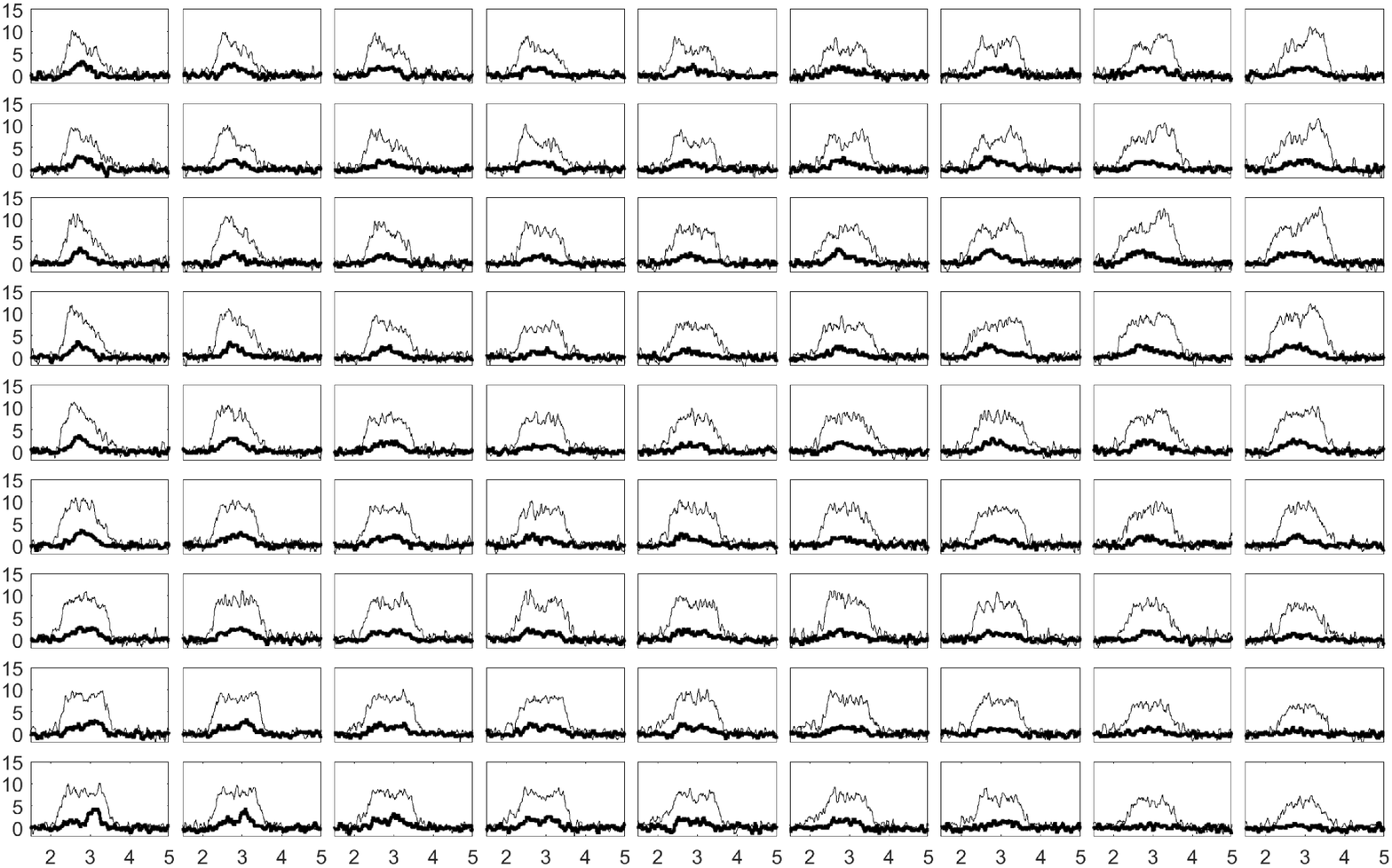}
      \caption{Same of \ref{fig:map1oso} but in MAP4.}
         \label{fig:map4oso}
   \end{figure*}

In Figs. \ref{fig:map1oso}, \ref{fig:map2oso}, \ref{fig:map3oso}, \ref{fig:map4oso} are shown the full maps observed with the Onsala telescope. Velocities are expressed in km s$^{-1}$ and the antenna temperatures in K. 

\section{H$_2$CO}\label{app:h2co}
\begin{figure*}
    \centering
            \includegraphics[scale=0.54,bb=40 5 1100 520,clip]{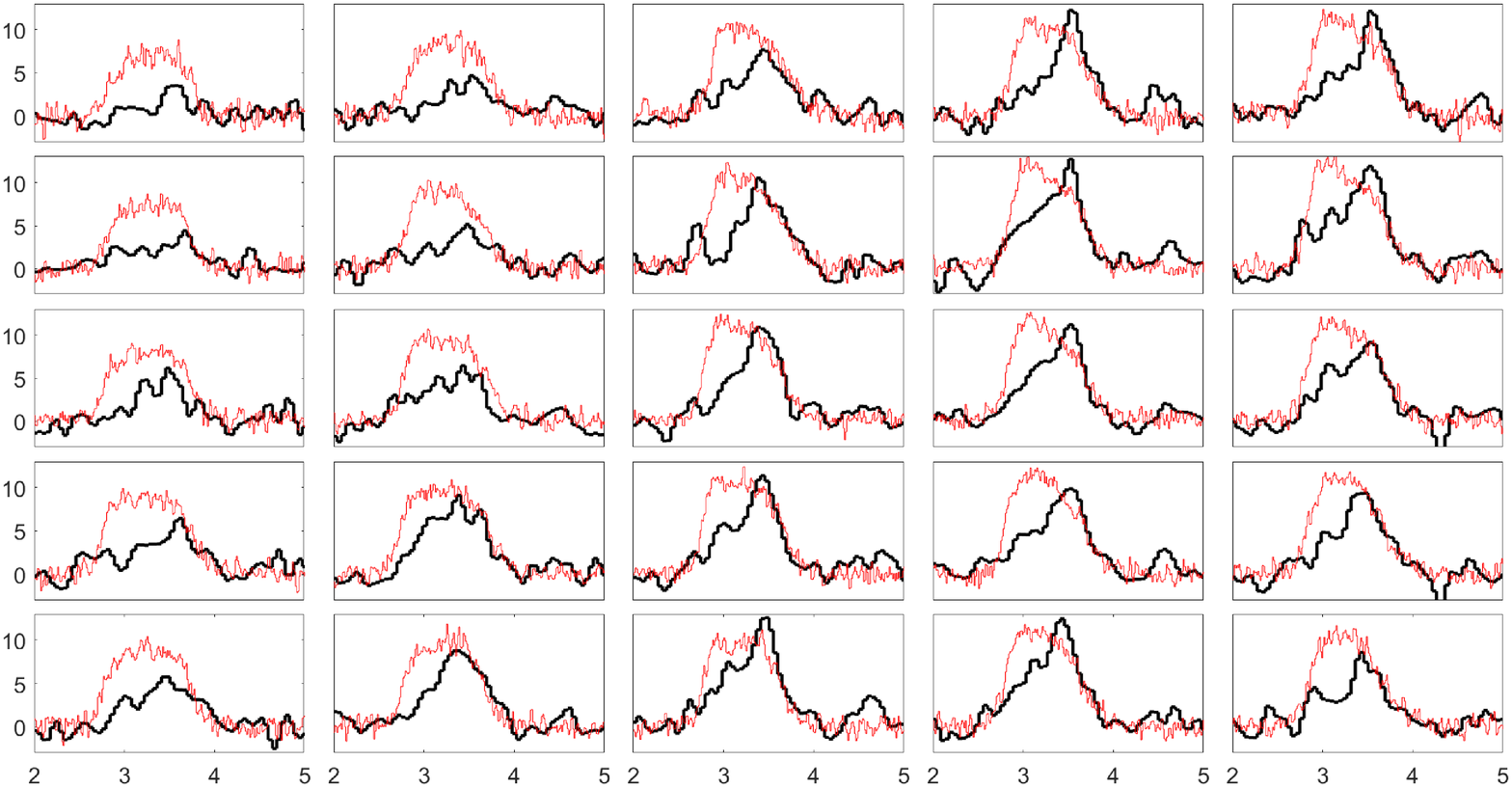}
      \caption{H$_2$CO observations in MAP1 (black thick line), compared with \element[][12]CO Onsala observations (red thin line) in the same map. H$_2$CO profiles are inverted and shown for scale to the \element[][12]{C}{O} peak temperature. Antenna temperature T$_A$ in K and v$_{LSR}$ in km s$^{-1}$.}
         \label{fig:map1_h2co_co}
   \end{figure*}

\begin{figure*}
    \centering
            \includegraphics[scale=0.54,bb=40 5 1100 520,clip]{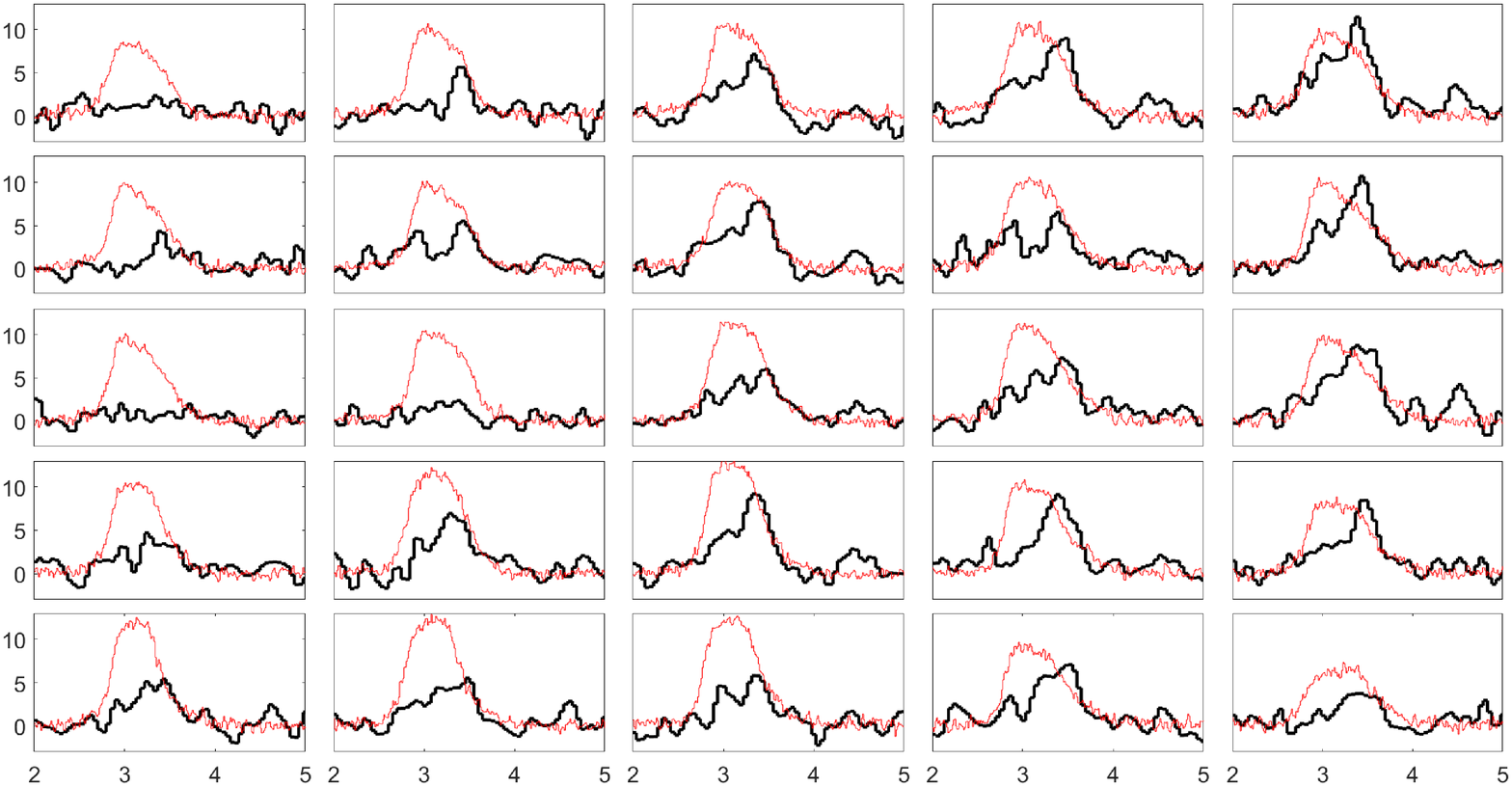}
      \caption{Same of \ref{fig:map1_h2co_co} but in MAP2.}
         \label{fig:map2_h2co_co}
   \end{figure*}

In Figs. \ref{fig:map1_h2co_co} and \ref{fig:map2_h2co_co} are shown the H$_2$CO observations. 

\section{Peripheral structures and ambient atomic filaments}\label{app:periph_struct}
\begin{figure*}
  \centering
  \includegraphics[scale=0.56,bb=-90 0 1100 620,clip]{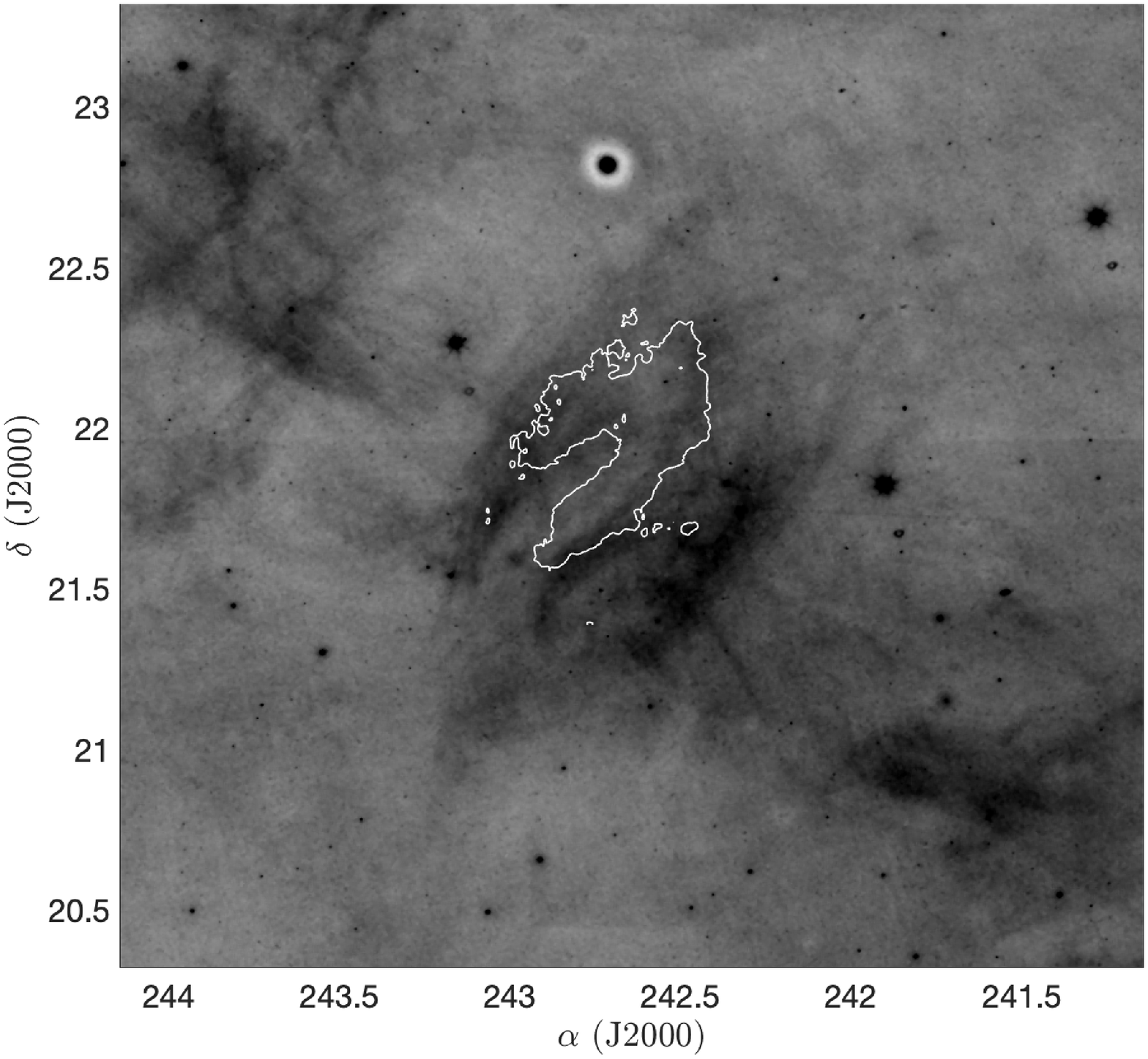}
     \caption{Gray scale plot of 12 $\mu$m image from \textit{WISE} to enhance the filamentary structure of cloud neighborhood. A single white contour at 1 K km s$^{-1}$ level from FCRAO \element[][12]{CO}{} is shown.}
        \label{fig:wise_filaments}
\end{figure*}

Some unusual low emissivity features are located around MBM 40, situated within the H I cocoon, and roughly parallel to the western ridge of MBM 40 on the side of the cloud roughly opposite the Galactic plane.  The structures are easily seen   
in the optical and near infrared (in Fig. \ref{fig:summary_obs} and in the {\it WISE} 12 $\mu$m data, Fig. \ref{fig:WISEPlanck}). They are also visible  in the {\it  Planck} 353 GHz  and the 100 $\mu$m {\it IRAS} images.  The features also appear in emission in the {\it WISE} 8$\mu$m PAH band, while MBM 40 does not. The larger field far infrared images demonstrate that these structures are filaments (see Fig. \ref{fig:wise_filaments}). The dust is associated with the diffuse 21 cm emission at v$_{LSR} < 3$ km s$^{-1}$, lower than that of the cloud, and the extended diffuse {\it IRAS} 100 $\mu$m emission.  
Since structures are discernible in the {\it Planck} and longest wavelength {\it IRAS} images, their emissivity must be thermal. Their different emission must, therefore, indicate changes in temperature and/or column density. Their lack of detection in the {\it WISE} short wavelength bands indicates that the dust is optically thin. There does not appear to be molecular gas associated with the structures but it is likely that these filaments have the same dynamical origin as the gas that which became MBM 40. Future studies may characterize the condensation from atomic to molecular gas in different environment, such as more compact diffuse cloud with different illumination or density knots seen in \ion{H}{i} observations. 

\end{appendix}

\end{document}